\documentclass[aps,prb,twocolumn,showpacs,eqseCNPum,superscriptaddress,floatfix,10pt]{revtex4-1}

\usepackage{amssymb}  
\usepackage{color,soul} 
\usepackage{graphicx}  
\usepackage[tight]{subfigure}  
\usepackage{mathrsfs}
\usepackage[pdftex,pdfusetitle,bookmarks=true,colorlinks=true,citecolor=blue,urlcolor=blue,linkcolor=magenta]{hyperref}
\usepackage{braket}
\usepackage{xfrac}
\usepackage{tikz}
\usepackage{url}
\usepackage{setspace}
\usepackage{enumerate}
\usepackage{bm}
\usepackage{epstopdf}
\usepackage{amsmath}
\usepackage{amsfonts}
\usepackage{mhchem} \usepackage{soul}

\newcommand{\blu}[1]{{\color{blue}#1}}

\begin{document}

\title{Hybrid Wannier Chern bands in magic angle twisted bilayer graphene and the quantized anomalous Hall effect}

\author{Kasra Hejazi}
\affiliation{Department of Physics, University of California Santa Barbara, Santa Barbara, California, 93106, USA}
\author{Xiao Chen}
\affiliation{Department of Physics and Center for Theory of Quantum Matter, University of Colorado, Boulder, CO 80309, USA}
\affiliation{Department of Physics, Boston College, Chestnut Hill, MA 02467, USA}
\author{Leon Balents}
\affiliation{Kavli Institute for Theoretical Physics, University of California Santa Barbara, CA 93106, USA}
\affiliation{Canadian Institute for Advanced Research, Toronto, Ontario, Canada}

\begin{abstract}
We propose a method for studying the strong interaction regimes in twisted bilayer graphene using hybrid Wannier functions, that are Wannier-like in one direction and Bloch-like in the other.   We focus on the active bands as given by the continuum model proposed by Bistritzer and MacDonald, and discuss the properties of corresponding hybrid Wannier functions. We then employ the method for a study of the fillings of $\pm 3$ electrons per moir\'e cell using the Hartree-Fock method. We discuss at length different regimes under which a quantized anomalous Hall effect is seen in these two fillings.
\end{abstract}

\maketitle

\section{Introduction}
Heterostructures containing moir\'e patterns due to incommensurations in multilayers containing graphene and other two dimensional crystals  have proven to be very tunable and promising platforms for observing interesting phases that are unprecedented in commensurate graphene systems\cite{cao2018correlated,cao2018unconventional,yankowitz2019tuning,lu2019superconductors,Sharpe605,serlin2020intrinsic}.
Twisted bilayer graphene (TBG) as the most prominent member has attracted much attention and also has given rise to numerous theoretical studies; however, still many of the different correlation induced phenomena in this system have eluded satisfactory theoretical understanding.

The most important theoretical discovery, probably, was the realization that a low energy theory, a continuum model (CM), could be effectively employed to study the single particle electronic properties of TBG at small twist angles\cite{bistritzer2011}; in fact, an analysis based on this CM resulted in the prediction of the possibility of strong correlation physics at the magic angle in the first place. Specifically in this CM, the smallness of the twist angle leads to an emergent periodicity in the system -- the so called \textit{moir\'e} lattice, which has a unit cell length growing like $\sim\frac{1}{\theta}$; such large periodicity in turn leads to formation of Bloch \textit{minibands}. Interestingly, around the magic angle, the bands closest to the charge neutrality point (CNP) show exceptional flatness and are well separated from other bands. Further including spin and valley degrees of freedom results in eight such bands in total. Since these bands are flat, the correlation between them can play an important role and give rise to interesting correlated phases and thus should be taken into account properly.
A possible theoretical approach to this end, is to consider an interacting model consisting of these \textit{active} bands only, treating the \textit{remote} bands as inert; we will be taking this route in this work and introduce a basis for the study of strong interactions.

Experimental observations of correlation induced insulating states have been reported in commensurate fillings of these active bands, along with superconducting behavior for fillings close to these commensurate values\cite{cao2018correlated,cao2018unconventional,yankowitz2019tuning,lu2019superconductors,Sharpe605,serlin2020intrinsic,saito2019decoupling,stepanov2019interplay}. Motivated by these experimental observations, here we pursue a theoretical model consisting of the subspace of the active bands only, in which electronic interactions are also projected onto this subspace; these interactions are local and thus working with local representations of the subspace spanned by active bands is desirable. However, as is well known, a faithful representation preserving manifest symmetries of the active bands using fully localized Wannier functions is difficult\cite{ahn2019failure,po2018origin}. Having this in mind, in this work, we work with Hybrid Wannier Functions (HWFs) which are Bloch-like in one direction and localized and Wannier-like in the other. Using this basis is a compromise between locality and symmetry/topology, noting that the wave functions are only localized in one direction, however, as is elaborated later, this ensures that important symmetries like valley and $C_2\mathcal{T}$ (the intravalley symmetry that protects the moir\'e Dirac points) remain manifest (when not broken at the non-interacting level). Furthermore, one ends up with a quasi-one-dimensional model, with local interactions in one direction, which can be suitable for numerical methods like DMRG\cite{kang2020non}.

As we show later, remarkably, full bands of these HWFs when maximally localized automatically exhibit a nonzero Chern number; this means that indeed a suitable collection of full bands of such states can display quantized anomalous Hall effect (QAHE), a phenomenon that has been reported in TBG\cite{Sharpe605,serlin2020intrinsic} at the filling factor of $\nu = +3$ (we define the filling factor $\nu$ to show the number of electrons per moir\'e cell measured from CNP). This makes the present \textit{maximally localized} HWFs a natural basis for a corresponding theoretical study. 
To analyze the effect of the interaction, which is evidently required for stabilizing a full band polarization in the HWF basis, we employ the self-consistent Hartree-Fock (HF) method at the two fillings $\nu = \pm 3$; these are the fillings where single fully occupied HWF bands of holes or electrons can be candidate many body states respectively.

We perform two separate studies of the effect of electron-electron interaction; first, we examine how the locality (in one direction only) of the HWFs makes full HWF bands advantageous for the interaction energy penalty when compared with other many body states at the same filling. Specifically, we check if full HWF bands turn out to be solutions of the HF equations when interaction is considered; this ensures that such HWF band polarized states have (at least local) minimal interaction energy compared with other candidate many body states. Second, we study the stability of similar many body states in a model obtained by projection of the full Hamiltonian onto the active bands. We present numerical results on the stability of QAHE in these two settings in a wide range of parameter choices of the models.

There have been other HF studies of the continuum model at various integer filling factors, with the analysis carried out completely using the basis of original Bloch states\cite{liu2019nematic,liu2019spontaneous,xie2020nature,bultinck2019ground}; in a subset of these works the remote bands are also kept in the analysis. The present study has the advantage of working directly with a faithful semi-localized representation of the active bands, while providing a continuous description of the QAHE with and without the $C_2 \mathcal{T}$ symmetry of TBG.  Moreover, in the present analysis, the QAHE appears naturally as polarized bands in the HWF basis and this could provide some more insight into the nature of the Chern bands responsible for this effect.
A comparison between these prior HF studies and our results is presented in Appendix \ref{app:comparison_hf}.

The paper is organized as follows: first, in Sec.~\ref{sec:model}, we demonstrate how the maximally localized HWFs are constructed and derive their topological properties. Then, in Sec.~\ref{sec:QAHE}, we present the HF study of the interacting model at the fillings $\pm3$, and the stability of QAHE by varying various parameters is examined. We conclude our results in Sec. \ref{sec:conclusion}.

\section{Hybrid Wannier functions}\label{sec:model}
We will be working with the continuum model introduced in Ref.~\onlinecite{bistritzer2011}. To take into account the two valleys, two parallel copies of the CM are considered;
in each copy, we will focus on the two \textit{active} bands, closest to CNP. Details of the non-interacting Hamiltonian are presented in Appendix \ref{app:model}. The CM has two free parameters in it: i) $\alpha \sim \frac{1}{\theta}w_{\text{AB}}$, which accounts for the collective effect of interlayer hopping $w_{\text{AB}}$ and the twist angle $\theta$, and ii) $\eta = \frac{w_{\text{AA}}}{w_{\text{AB}}}$, the ratio of the interlayer tunneling strength in AA and AB regions of the moir\'e lattice,  which encodes how much corrugation is present in the system.
 We will also consider adding a sublattice symmetry breaking term $\Delta \, \sigma^z$ to the non-interacting Hamiltonian, where the Pauli matrix $\sigma^z$ is used to address sublattice degrees of freedom; this could account for the effect of aligned hexagonal Boron Nitride (hBN) substrates on the two sides of the TBG sample. \footnote{More relevant to experiments is a setup with different subalttice potentials on the two layers, but here for simplicity we take the potential to be identical on both layers.}
 We will also be using an approximation\cite{hejazi2019multiple,song2019all} which renders a particle-hole symmetry to the CM; this approximation becomes better at small angles, see Appendix \ref{app:model} for details. 

\begin{figure*}[!t]
	\centering
	\subfigure[]{\label{fig:real_space_lattice_BZ}\includegraphics[width=0.485\textwidth]{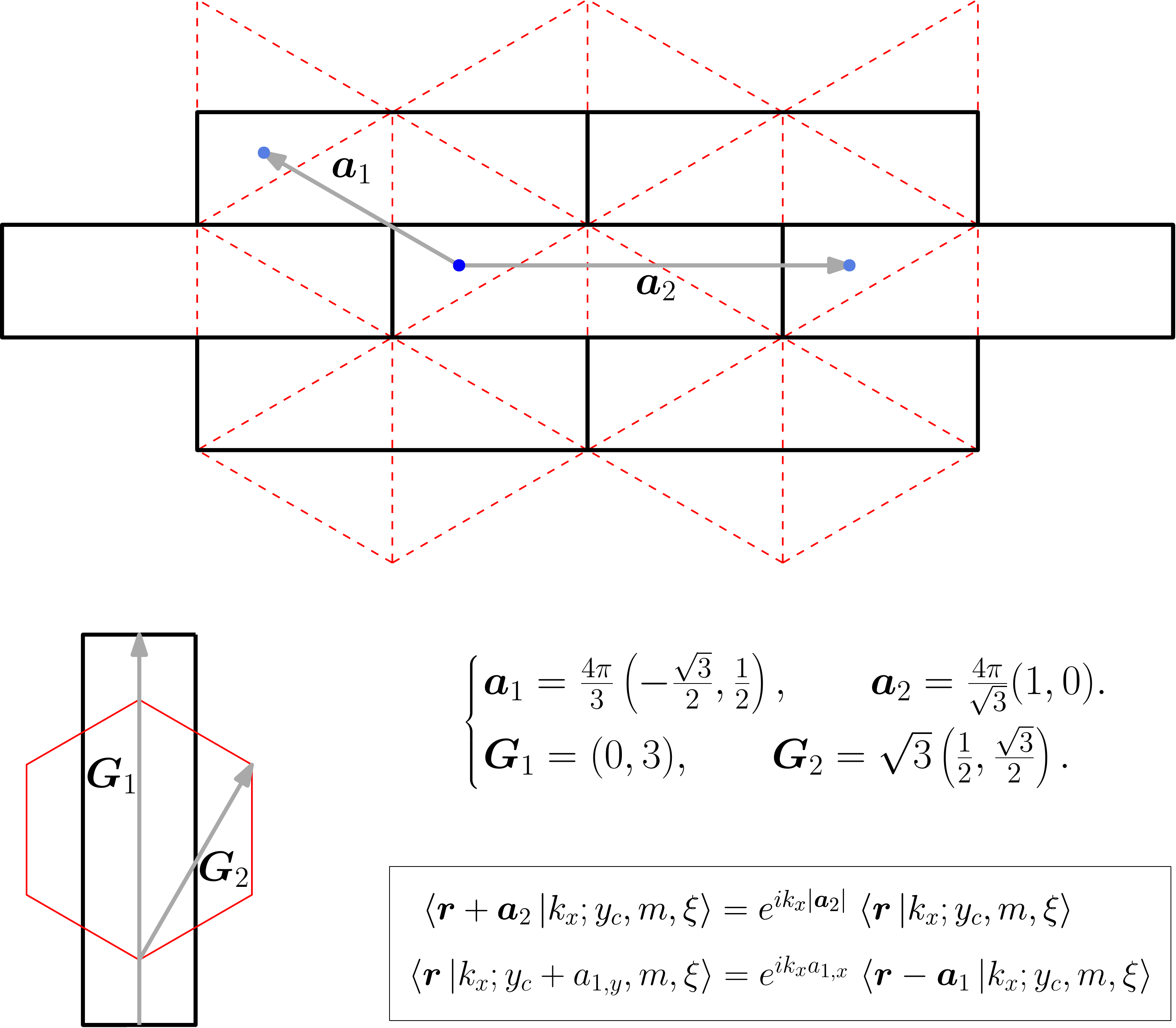}}
	\hspace{1.5cm}
	\subfigure[]{\label{fig:wilson_lines}
	\includegraphics[width=0.32\textwidth]{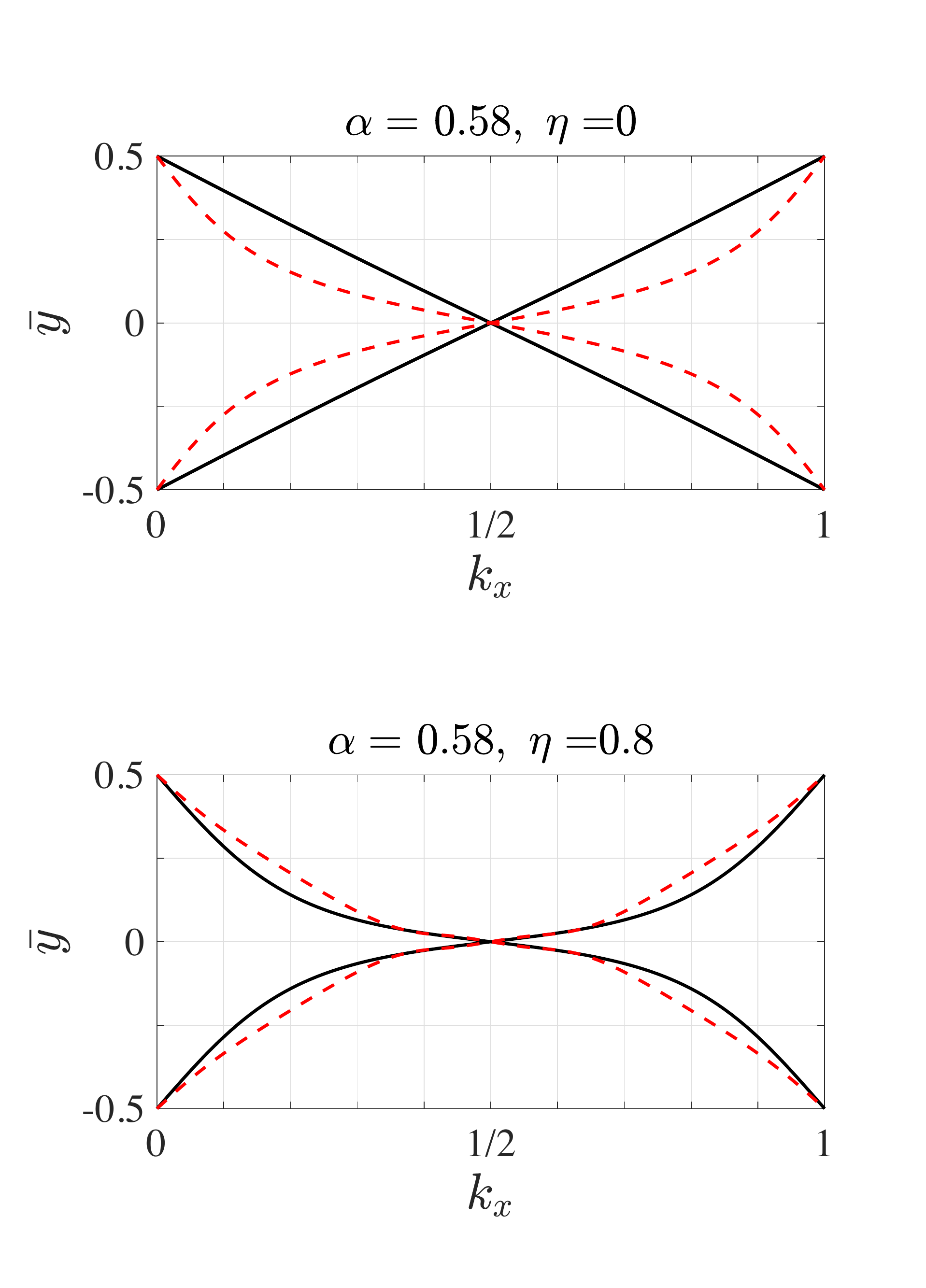}}
	\caption{\footnotesize 
	(a) The moir\'e lattice in real space and the corresponding BZ. A rectangular BZ is chosen $-\frac{\sqrt{3}}{4}<k_x<\frac{\sqrt{3}}{4}, \ -\frac32 < k_y < \frac32$; note that this is contrary to the usual hexagonal choice so that the top and bottom of the BZ are identified, note that this is crucial for the usual properties of the one-dimensional Wannier transform in the $y$ direction to hold. The equations governing the translational properties of the HWFs are also presented.
	   (b) WCC positions (solid black lines) and single band Berry phases in the original Bloch bases (dashed red lines) of the two active bands. One of the plots corresponds to the chiral model and the other to the physical value of $\eta$. A small sublattice potential is added, $\Delta = 0.19 \text{meV}$. The configuration of the dashed lines and the solid lines mean that the two bands carry $+1$ and $-1$ Chern numbers in the original Bloch representation and the parallel transport representation respectively. This is a robust feature present in a wide range of parameter choices. Note that $k_x$ is rescaled and instead of plotting the interval $[-0.5,0.5)$, equivalently $[0,1)$ is drawn.
	   } 
	\label{fig:fig_1}
\end{figure*}

Equipped with the full non-interacting content of the model, one can find the Bloch states lying in the middle two active bands for each valley. We take the active bands to be well separated from the remote bands, and thus develop an active-bands-only model. Following the notation and methods introduced in Refs.~\onlinecite{marzari1997maximally,marzari2012maximally} we will Wannier transform properly chosen Bloch states in only one direction to obtain the \textit{maximally localized} HWF basis as follows:
	\begin{equation}\label{eq:HWF_def}
		\begin{aligned}
			\left|  k_x ; y_c , m , \xi \right \rangle &= \frac{1}{N_y} \sum_{k_y} e^{-i k_y y_c}  \left|  \tilde{\psi}_{ \bm{k} ; m , \xi} \right \rangle, \\
			\left| \tilde{\psi}_{ \bm{k} ; m , \xi}\right \rangle &=  \sum_{y_c} e^{i k_y y_c}  \left|   k_x ; y_c , m , \xi  \right \rangle,
		\end{aligned}
	\end{equation}
where $\left|  k_x ; y_c , m , \xi \right \rangle$ stands for a Hybrid Wannier state, with the indices  $y_c , m , \xi$ denoting the real space position in the localized direction, the band (orbital), and the valley respectively.  The states on the right hand side are linear combinations of the Bloch eigenstates of the non-interacting Hamiltonian at each $\bm{k}$: 
	\begin{equation}\label{eq:bloch_redefinition}
		\left|  \tilde{\psi}_{ \bm{k} ; m , \xi} \right \rangle = \sum_n \left|  \psi_{\bm{k} ; n , \xi} \right \rangle U_{nm}^{\bm{k},\xi}.
	\end{equation}
The unitary (in the band basis) matrices $U$ are chosen at each $\bm{k}$ to ensure that maximal localization is achieved in the $y$ direction ultimately and the procedure is detailed below.
Here, a rectangular BZ is chosen as shown in Fig.~\ref{fig:real_space_lattice_BZ} so that the $k_y$ sum needed for a one dimensional Wannier transform in \eqref{eq:HWF_def} is performed at each $k_x$. The spin index trivially doubles all manipulations here and thus is suppressed. With the above convention, the allowed values of $y_c$ form a one dimensional lattice with lattice spacing equal to half a moir\'e length ($\frac12 a_M=a_{1,y}$), i.e.~$y_c = j \frac{a_M}{2}$ where $j$ is an integer. Note that we take this lattice to be identical for different values of $k_x$, and so the above $y_c$ values are different from but close to the actual locations of Wannier charge centers (WCC) of HWF states (see below for more information). The transformation of the HWFs under moir\'e lattice translations is depicted in Fig.~\ref{fig:real_space_lattice_BZ}.

In order to obtain maximal localization, one needs to choose the matrices $U$ in \eqref{eq:bloch_redefinition} properly: to this end, we will use the procedure  discussed in Ref.~\onlinecite{marzari1997maximally} to form the \textit{parallel transport} basis for the Bloch functions, an approach that is suitable for maximal localization of one dimensional Wannier functions, and in the present study should be carried out for each strip with a definite $k_x$ separately. We will use a discretization which will lead to a Bloch momenta lattice with lattice spacings $b_x , b_y$, and with $N_x , N_y$ total points along the two directions.
According to this prescription, at each $\bm{k}$, the overlap matrices 
\begin{equation}
    \mathcal{M}_{mn}^{k_x,k_y,\xi} = \left\langle u_{ k_x , k_y ; m , \xi} \right.\left|  u_{ k_x , k_y+b_y ; n , \xi} \right \rangle, 
\end{equation}
are calculated, where as usual $\left|  u_{\bm{k} ; n , \xi} \right\rangle$ shows the unit cell periodic part of an original Bloch function; notice that there is a small displacement in the $k_y$ direction in the ket state. Next, redefinitions of Bloch functions are made as shown in \eqref{eq:bloch_redefinition}, with $U$ matrices chosen in a way that the updated $\mathcal{M}$ matrix for all $k_x,k_y,\xi$ attains a form as $K^{k_x,k_y,\xi} \ \gamma^{k_x,\xi}$, where $K$ is Hermitian and $\gamma$ is diagonal, unitary and independent of $k_y$. This, as discussed in Appendix \ref{app:max_local_HWF}, ensures maximal localization in the $y$ direction.

The path ordered product of all $\mathcal{M}$ matrices along a strip with a given $k_x$ defines its Wilson loop, whose eigenvalues are invariant under a $\bm{k}$ dependent basis change such as the one in \eqref{eq:bloch_redefinition}.
One can show that the $K$ matrices as defined above are equal to the identity matrix to first order in $b_y$ and thus, the eigenvalues of each $\gamma^{k_x,\xi}$ matrix above are directly related to the Wannier charge center positions, i.e.~Wilson loop eigenvalues, in the strip given by $k_x$.
Using this fact, WCCs of HWFs as functions of $k_x$ could be found with examples drawn  in Fig.~\ref{fig:wilson_lines}. It could be seen by inspection that, regardless of the set of parameters chosen, there is a $+ 1$ winding and a $-1$ winding of the WCCs for the two HWF bands as $k_x$ traverses the BZ.\cite{liu2019pseudo,song2019all}
Noting, based on the above observations, that in the parallel transport basis, the single band Berry phases along each strip with a given $k_x$ are equal to the WCC values,
leads us to an important implication for the parallel transport basis:
given how WCCs behave as functions of $k_x$ shown in Fig. \ref{fig:wilson_lines}, the two Bloch bands in the parallel transport basis have Chern numbers $+1$ and $-1$.
This, in other words, means that a fully filled band of maximally localized HWFs exhibits a quantized Hall response. As a result, when addressing the \textit{maximally localized} HWFs, the terms band, orbital and Chern number could be used interchangeably.
 
In some special cases,
the parallel transport basis can be found explicitly. For instance, when $\Delta = 0$, there is a $C_2 \mathcal{T} = \sigma^x \mathcal{K}$ symmetry of the Hamiltonian, where $\mathcal{K}$ is the complex conjugation operator; as shown in Appendix \ref{app:max_local_HWF},
the combinations $\frac{e^{ \pm i \phi_{\bm{k},\xi}}}{\sqrt{2}}   \left( \left| \psi_{\bm{k};1,\xi} \right\rangle   \pm i \left| \psi_{\bm{k};2,\xi} \right\rangle \right) $, 
with the phases $\phi_{\bm{k},\xi}$ appropriately chosen, form the parallel transport Bloch basis at $\bm{k}$, where states $\left| \psi_{\bm{k};m,\xi} \right\rangle$ show $C_2 \mathcal{T}$ symmetric Bloch eigenstates. In particular, if one now sets $\eta = 0$ to obtain the chiral limit, since the two $C_2\mathcal{T}$ symmetric bands are related\cite{tarnopolsky2019origin} by $\left| \psi_{\bm{k};1,\xi} \right\rangle   = i \sigma^z \left| \psi_{\bm{k};2,\xi} \right\rangle $, the parallel transport basis consists of sublattice polarized states;
remarkably, even with $\Delta \neq 0$ while keeping $\eta = 0$ this result holds, i.e.~the parallel transport basis consists of sublattice polarized states. 
By numerical inspection, one can show that each of the two bands in the parallel transport basis is more concentrated on one of the sublattices to a high degree  in a one-to-one fashion, even away from the chiral limit. It is worthwhile to mention that the $U(4) \times U(4)$ symmetry discussed in Ref.~\onlinecite{bultinck2019ground} (which states that the interaction term of the Hamiltonian is invariant under rotations of the bands with equal Chern numbers into each other) could be seen readily in the above construction of the parallel transport basis. 
This along with other symmetries of the CM as seen in the HWF basis are discussed in length in Appendix \ref{app:hamiltonian_hwf}.

The HWF basis naturally defines the problem in the geometry of a cylinder. The HWFs form ring shaped wires around the cylinder, since these wave functions are localized in one direction and extended in the other.
Each wire is identified with a $y_c$, and is composed of states with different values for their $k_x$, band number, valley number and spin (see Eq.~\eqref{eq:HWF_def}). At the non-interacting level, hopping occurs between states in separate wires if they have identical $k_x$, valley number and spin (See Appendix \ref{app:hamiltonian_hwf} for details). This hopping decays as the distance between wires along the cylinder is increased.
Based on this HWF construction, in the next section we will present a HF study of a model consisting of active bands only with a total Hamiltonian of the form:
\begin{equation}\label{eq:hamiltonian_terms}
\begin{aligned}
	H &= H_{\text{kin}} + H_{\text{int}}   + H_{\text{MF},0}.
\end{aligned}
\end{equation} 
$H_{\text{kin}}$ contains the single particle terms in the Hamiltonian induced by the CM, i.e.~the hoppings between different wires as mentioned above.   The remaining two terms represent effects of interactions: they are both proportional to $e^2/\epsilon$, where $e$ is the electron charge and $\epsilon$ is the dielectric constant, and thus vanish in the non-interacting limit.  $H_{\text{MF},0}$, which is  quadratic in fermion operators, is responsible for two separate effects: it takes the effect of filled remote bands into account at a mean field level and it also serves to avoid a double counting of HF terms that are already taken into account in $H_{\text{kin}}$\cite{liu2019nematic,bultinck2019ground} (see the discussion at beginning of the next section for more details).
Turning to the interaction term $H_{\rm int}$, we have chosen the electron-electron potential to have a screened coulomb form as $V_{\text{int}}(\bm{r}) = \frac{e^2}{4\pi \epsilon } \frac{e^{- \left| \bm{r} \right| / \ell_\xi}}{\left|\bm{r} \right|}$, which is further projected onto the active bands. The interaction retains its \textit{normal-ordered} density-density form with respect to spin, sublattice, layer and valley indices (more details are presented in Appendix \ref{app:hamiltonian_hwf}). 
Note that due to the locality of HWFs, the electron-electron interaction between the wires drops as the distance between them is increased, and thus the total Hamiltonian is local in the direction along the cylinder.

We conclude this section by some remarks regarding the parameter values and conventions used: via dividing the energies and lengths by $\hbar v_F k_{\theta}$ and $\frac1{k_\theta}$ respectively, we have made them dimensionless, where $k_\theta = \frac{4\pi}{3} \frac{1}{a_M}$.
In this notation, we define the dimensionless interaction strength parameter $g_{\text{int}} = \frac{e^2}{2 \epsilon } \frac{1}{\hbar v_F k_\theta^2 \mathcal{A}}$, where $\mathcal{A}$ is the area of a moir\'e unit cell. Numerically $g_{\text{int}} = 1.01 \frac{\epsilon_0}{\epsilon}$, and thus, a choice of $\epsilon = 7 \epsilon_0$ results in $g_{\text{int}} = 0.14$ as an example.

The model introduced above comprises bands (in the parallel transport basis) that carry nonzero \blu{$\pm 1$} Chern numbers; thus a quantized Hall signal can be observed at integer filling factors if with some interaction induced effect, a suitable valley and band polarization in the system occurs. As a result, it is natural to utilize the present model to study the physics of quantized anomalous Hall effect (QAHE) seen in some samples of twisted bilayer graphene\cite{Sharpe605,serlin2020intrinsic}. We will do so in the following section for the two fillings $\nu = \pm3$.

\section{Quantized anomalous Hall effect in twisted bilayer graphene}\label{sec:QAHE}

In this section, we present two separate HF studies in which different choices of $H_{\rm MF,0}$ are utilized.  We focus on the filling $\nu=\pm 3$ and  explore the stability of QAHE phase in these two different schemes. Before we go into the detail, we first discuss how the HF procedure is carried out in general.

The HF procedure is implemented as follows: we fix the filling and seek a Slater determinant many body state, composed of single particle states $\left|\psi_l \right\rangle = \sum_{\alpha} \psi_{l,\alpha} \left|\alpha \right. \rangle$, that minimizes the expectation value of the Hamiltonian \eqref{eq:hamiltonian_terms}, where $\alpha,\beta,\ldots$ denote the HWF basis indices $k_x,y,\xi,m,s$ (the states $\left| \alpha \right\rangle$ will be normalized in this section). One seeks $\left|\psi_l \right\rangle$ states by transforming the Hamiltonian \eqref{eq:hamiltonian_terms} written in the form
\begin{equation}
\begin{aligned}
	H &= \sum_{\alpha\beta} H_{0,\alpha\beta} \ c^\dagger_\alpha c^{\phantom{\dagger}}_{\beta} \\
	&+\frac12 \sum_{\alpha\beta \, \alpha'\beta'} V_{\alpha,\beta,\beta'\alpha'}   \  c^\dagger_\alpha c^\dagger_\beta c^{\phantom{\dagger}}_{\beta'} c^{\phantom{\dagger}}_{\alpha'} ,
\end{aligned}	
\end{equation}
into a single particle \textit{HF Hamiltonian}, wherein the interaction term is transformed into
\begin{equation}
\begin{aligned}
	H^{\text{HF}}_{\text{int}} &= \sum_{k_1k_2,aa'bb'}  c^{\dagger}_{k_2b}  c^{\phantom{\dagger} }_{k_2b'}  \,  P(k_1)_{aa'} \\
	&   \left[ V_{k_1a,k_2b,k_2b',k_1a'} -  V_{k_1a,k_2b,k_1a',k_2b'} \right].
\end{aligned}	
\end{equation}
In the above, $a,b,\ldots$ (contrary to $\alpha,\beta,\ldots$) show the HWF indices except $k_x$. Notice that we have dropped the $x$ subscript from $k_x$ and will do so from now on; the $k$ dependent matrices $P$ have the form $P(k)_{aa'} = \sum_{l} \psi_{l, k a}^* \, \psi^{\phantom{*}}_{l, k a'}$. It has, furthermore, been assumed that the translational symmetry around the cylinder is not broken. 

The above HF Hamiltonian depends on its own eigenstates and thus we aim to obtain them iteratively: starting from a well chosen initial many body state, at each iteration step, $P$ matrices are updated using the eigenstates found in the previous step; a $\nu$ dependent number of these eigenstates with lowest eigenvalues participate in forming the $P$ matrices. The resulting HF Hamiltonian is then diagonalized to yield the updated set of eigenvalues and eigenstates. This procedure is continued until convergence is achieved. We obtain the sought HF many body state as a slater determinant of the converged eigenstates with lowest HF eigenvalues. Moreover, the nearby eigenvalues above and below the ``Fermi energy'' could be used to give estimates of the actual energies needed for adding or removing an electron at this filling (Koopmans' theorem\cite{ostlund1982modern}) \footnote{Koopmans' theorem yields the change in energy if an electron or a hole is added to an $N$-particle HF state, while assuming that the $N$ electrons' states are unaltered. This is not necessarily a good approximation even in the HF approach. However, we mostly use a gap in the HF eigenvalues to determine whether QAHE is stabilized or not as detailed in the main text.}.

The two approaches mentioned at the beginning of this section are taken into account by two different choices for $H_{\text{MF},0}$ in the Hamiltonian \eqref{eq:hamiltonian_terms}. In the first study, Sec.~\ref{sec:fs}, we examine the motivation with which the HWF basis was introduced:  the interaction energy of different many body states are compared with $H_{\text{MF},0} = 0$. 
In particular, the energy of the state that is described as a full band of electrons ($\nu=+3$) or holes ($\nu=-3$) in the HWF basis is compared with other HF many body states. 
Note that this choice of $H_{\text{MF},0} = 0$ results in a competition between the interaction energies and the band structure energies as given by the CM; the latter, which could also be viewed as the hopping term in the HWF basis, is kept in the analysis so that one attains a measure for defining strong and weak interaction regimes. 

In the second study, in Sec.~\ref{sec:ss}, on the other hand, we take\footnote{
This choice is similar to the one in Ref.~\onlinecite{liu2019nematic}.} 
\begin{equation}\label{eq:H_MF_second_approach}
    H_{\text{MF},0} = - \sum_{\alpha\alpha'} \left[  \sum^\prime_{\beta\beta'} \left( V_{\alpha,\beta,\beta',\alpha'} - V_{\alpha,\beta,\alpha',\beta'}  \right) \right] \, c_{\alpha}^{\dagger} c_{\alpha'}^{\phantom{\dagger}}
\end{equation}
where the $\alpha,\alpha'$ summation is done over {\it all} states in the active bands, but the {\it partial} summation over $\beta,\beta'$ (indicated by the prime on the sum) ranges {\it only} over those states in the active bands that are below the CNP of the CM. Note that the latter states when written in terms of the HWF basis will not be band diagonal. By taking $ H_{\text{MF},0}$ to have the form in Eq.~\eqref{eq:H_MF_second_approach}, we are taking two separate effects into account: first, a mean field potential induced by the filled remote bands. The second effect, instead, has to do with the fact that within HF, the electron/hole dispersion will only agree (at best) at one filling with the dispersion given by the term $H_{\text{kin}}$. We take that point to be the CNP of the CM bands in the second study, i.e.~we assume that the CNP dispersion given by the CM, describing single electron or single hole excitation energies on top of the CNP, is unaltered by HF (see Appendix \ref{app:hamiltonian_hwf} for discussion). In order for this to be true, a HF effect of all filled bands (including remote and active bands) at the CNP is subtracted. The combination of these two effects results in a cancellation of the mean field effect of the filled remote bands and thus one ends up with the form in \eqref{eq:H_MF_second_approach} with only the mean field effect of active filled bands subtracted.

In the next two subsections, we present our numerical results corresponding to these two studies.

\subsection{First study}
\label{sec:fs}

In this subsection, we consider a model in which $H_{\text{MF,}0} = 0$, wherein a competition between electron-electron interactions and the noninteracting hopping in the HWF basis enables us to tune the model into and out of the strong coupling regime.
Previous studies, working on generic models similar to the one used in this subsection, have shown analytically that in strong coupling limits, valley polarization in these two filling factors is expected\cite{alavirad2019ferromagnetism,repellin2019ferromagnetism}. Here, we present a more thorough HF study of the Hamiltonian, trying to identify different regimes in which QAHE could be achieved.

In a given setting, we say that the QAHE is stabilized through HF if two requirements are met: i) if we initialize the HF iterative process with a fully spin-valley-band-polarized state, the HF iterations lead to a final HF state that is only achieved through smooth deformation of the spin-valley-band-polarized state (see below for further discussion of this notion of smooth deformation), and ii) the final HF solution properties in large enough systems do not change considerably as the system size is varied. 
 
This means that the QAHE state is at least a minimum of energy; we have also tried perturbing the final HF state in different ways to examine the stability of the HF solutions and we have observed that the final many body states generally show a high level of stability (see the discussion right above Sec.~\ref{sec:ss} for more on other possible HF solutions). 
In each setting we start with strong interactions first and see if the QAHE state is stabilized, and then continue to lower the interaction strength. We will, furthermore, use periodic boundary conditions along the cylinder. For the numerical results presented in this paper, the system is chosen to have $N_x=N_y=20$.

\begin{figure}[!t]
	\centering
	\subfigure[]{\label{fig:fillings_eta_0_Delta_0}\includegraphics[height=0.24\textwidth]{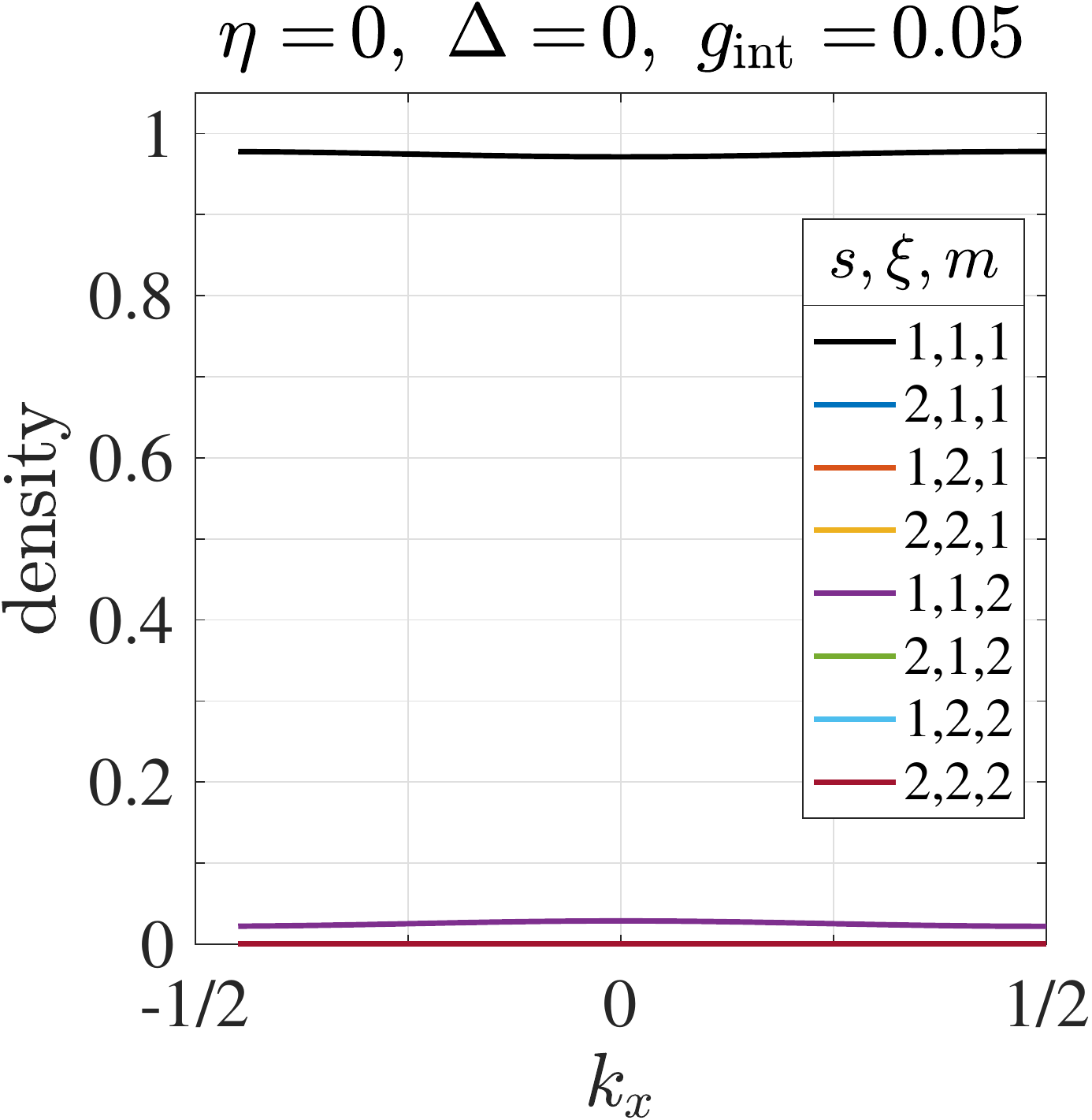}}
	\subfigure[]{\label{fig:energies_eta_0_Delta_0}\includegraphics[height=0.24\textwidth]{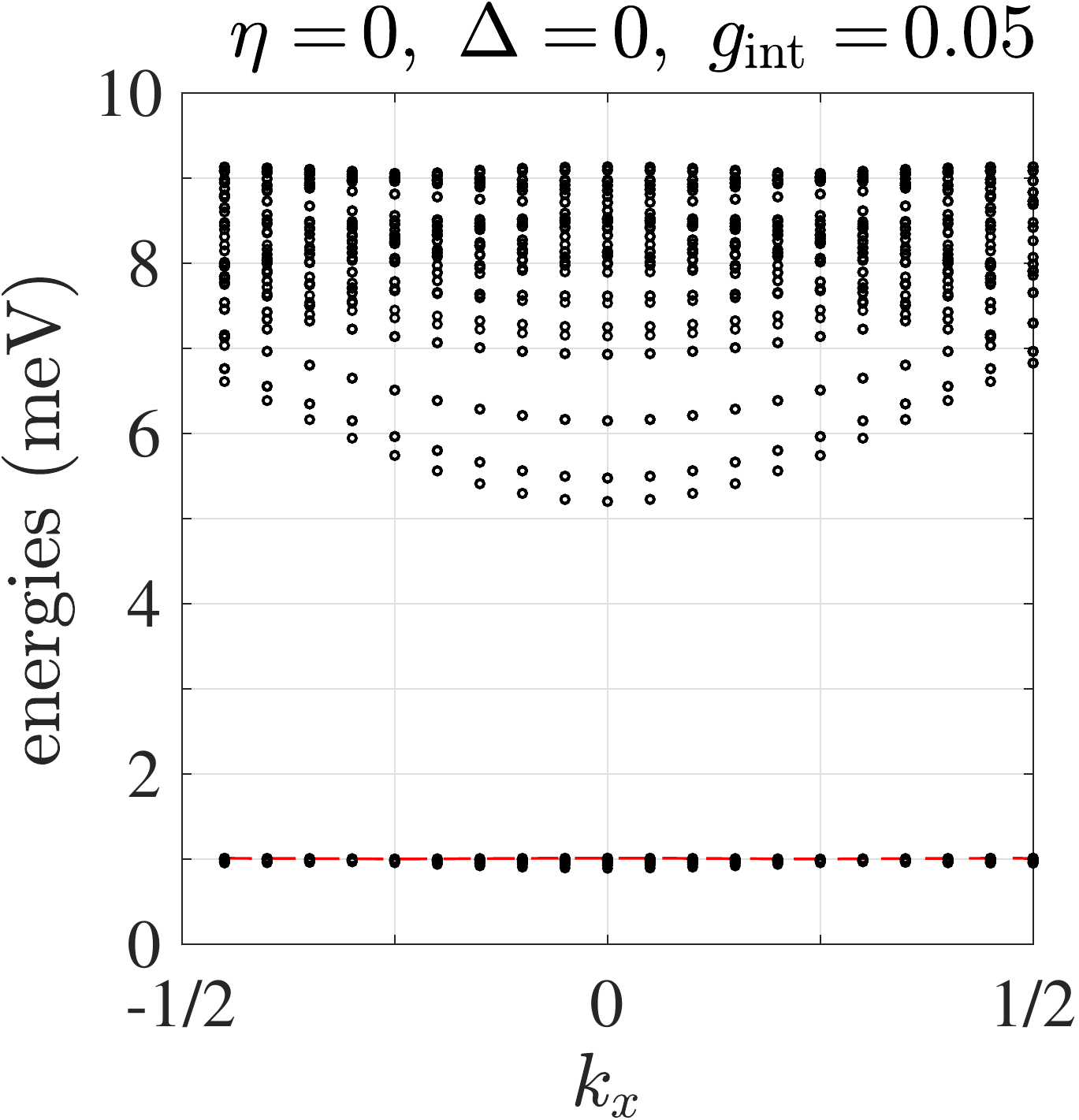}}
	\subfigure[]{\label{fig:energies_eta_0_Delta_01}\includegraphics[height=0.24\textwidth]{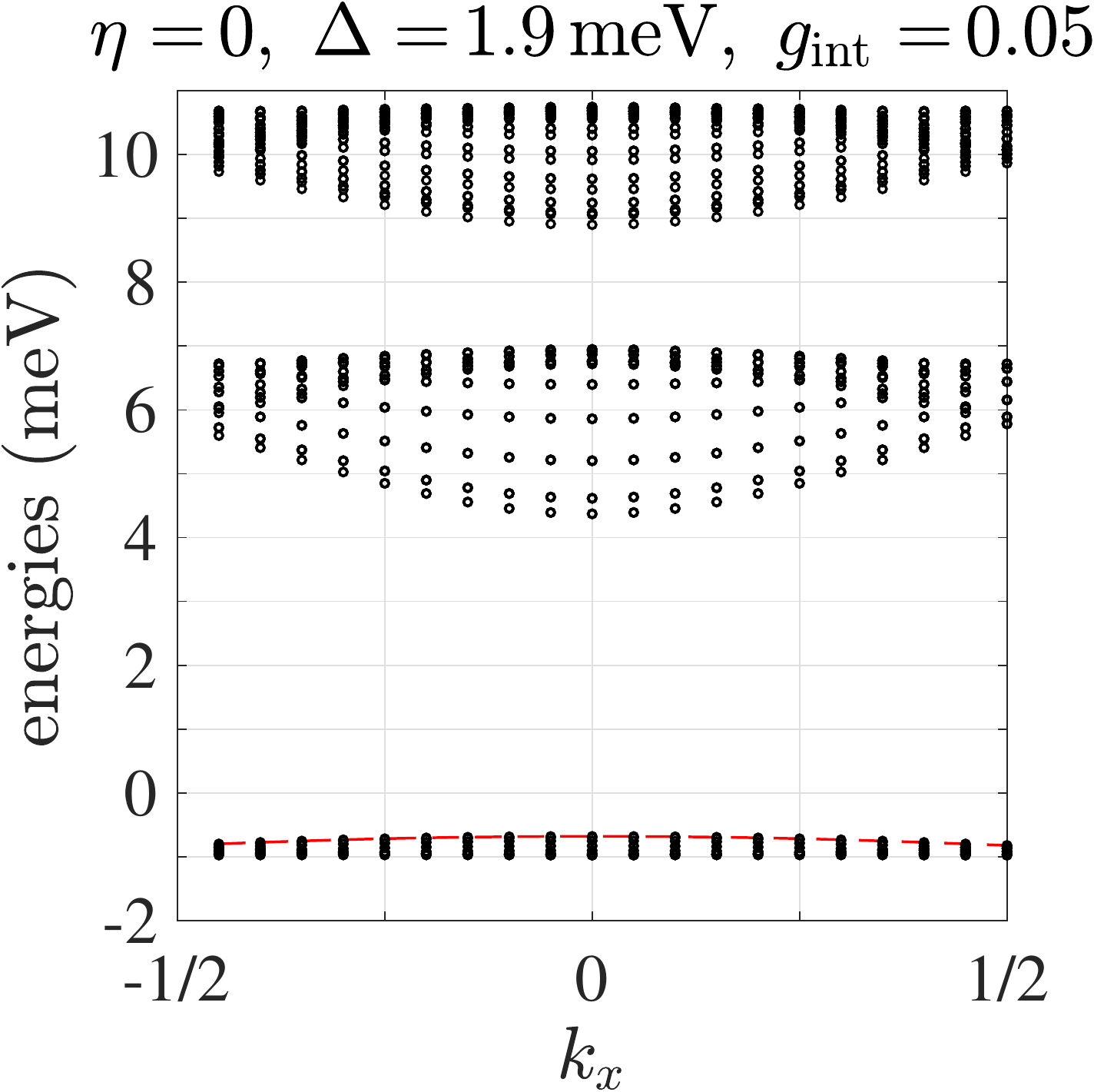}}
\hspace{-.3cm}
	\subfigure[]{\label{fig:gaps_eta_0}\includegraphics[height=0.225\textwidth]{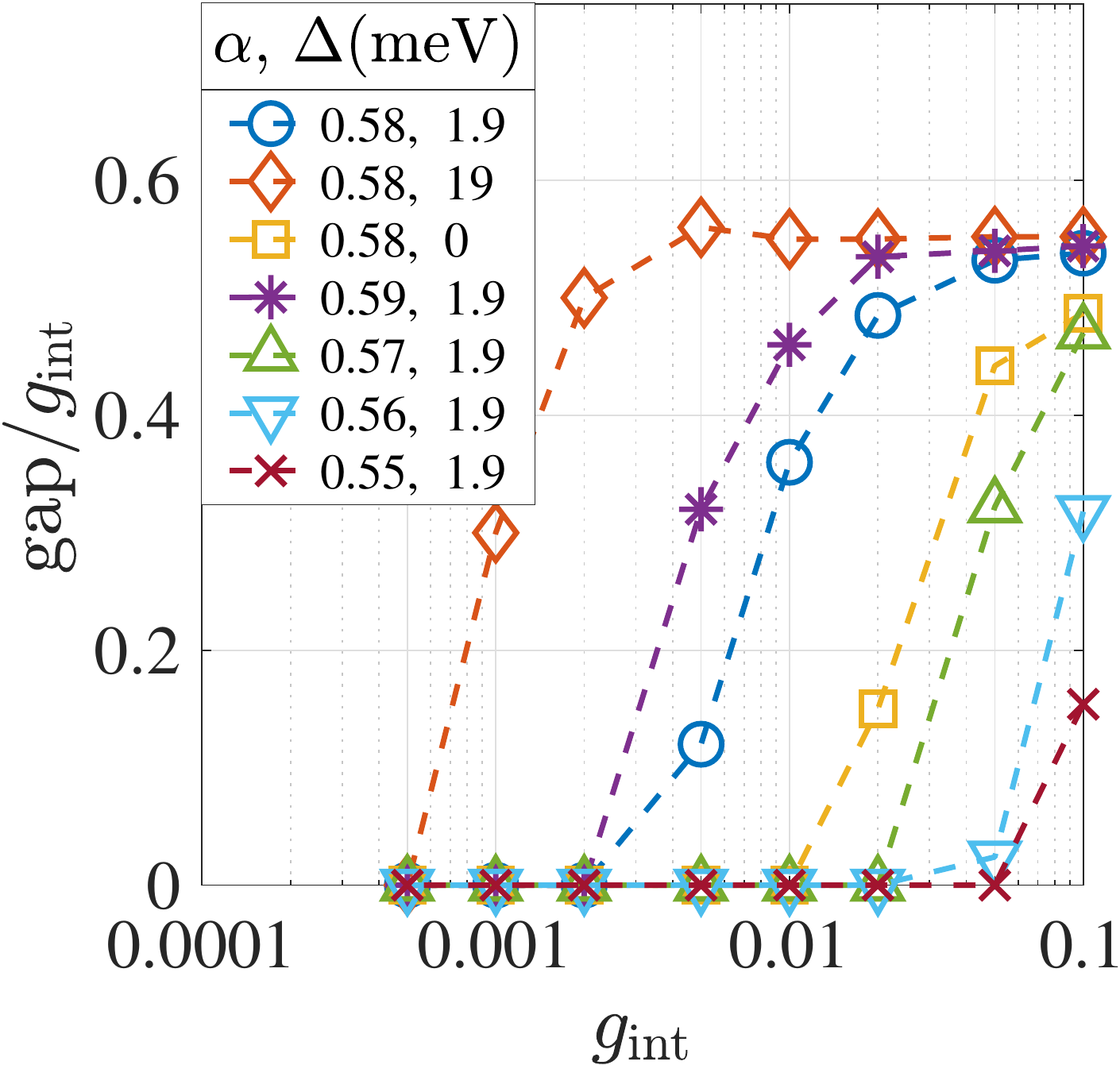}}	
	\caption{\footnotesize The filling $\nu=-3$, $\eta = 0$, and $\ell_\xi = 0.17 a_M$ plots within the first study, the angles are chosen around the magic value ($\alpha_0 = 0.586$ corresponding to $\theta=1.05^\circ$) (a) Density of electrons (number of electron per unit length of the cylinder) vs $k_x$ (momentum across the cylinder) for different flavors where $s,\xi,m$ stand for spin, valley, and band. $\alpha = 0.58$, $\Delta = 0$ and furthermore $g_{\text{int}} = 0.05$ have been chosen here, this value for the latter corresponds to being close to strong interaction limit since the band width is very small. Almost full polarization is seen here. (b) The HF eigenvalues (energies), with the same parameters as described in (a). The Fermi surface is shown with the dashed red line, one can observe a HF gap which will be used as a criterion for determining whether QAHE has been stabilized under HF iterations or not. (c) The same plot as in (b) with $\Delta=1.9 \text{meV}$. One can observe a second gap that is formed above the interacting one, due to the relatively large $\Delta$ chosen; it  separates the states belonging to the opposite sublattices. (d) The HF gap divided by $g_{\text{int}}$ as a function of $g_{\text{int}}$ for several parameter choices. 
	It can be inferred that approaching the magic angle and making $\Delta$ larger makes the QAHE more HF stable.
	}
	\label{fig:eta_0_plots_first_study}
\end{figure}

\subsubsection{ $\eta = 0$, short range interaction}
We now start to present our numerical HF results. We construct the basis of HWFs by forming the parallel transport basis on a finite lattice in $k$-space as discussed in Appendix \ref{app:max_local_HWF}, and then make a Wannier transform along $y$ for each $k_x$.

The chiral model, i.e.~when $\eta=0$, is first considered, in which absolutely flat bands are achieved at the magic angle. We start with the small $\ell_\xi$ limit so that the electron-electron interaction $V_{\text{int}} (\bm{r})$ is very short ranged. More realistic longer range interaction is considered later. Moreover, we also take the twist angle different from but close to the magic value so that the bands exhibit a nonzero small width.

In the first setting outlined above, or concretely with the choices $\eta = 0$ and $\ell_\xi = 0.17 a_M$, numerical analysis shows that the QAHE is generically stabilized at $\nu=-3$ at large interaction strength, see Fig.~\ref{fig:eta_0_plots_first_study}(a,b), where the density of different flavors along with HF eigenvalues (energies) are shown for an instance where the interaction plays the dominant role. Note that because of the nature of the HWF basis, this is a $C_2\mathcal{T}$ broken many body state, despite the fact that this symmetry is present at the noninteracting level. We define a HF gap as the lowest unoccupied HF eigenvalue minus the highest occupied eigenvalue, this quantity when divided by the interaction strength $g_{\text{int}}$ serves as a good qualitative measure of whether and to what extent the polarized state is stabilized under HF.

Note that in Fig.~\ref{fig:eta_0_plots_first_study}(a), although the many body state has components in both HWF bands in a single valley and spin sector, and the two bands have opposite Chern numbers, the Hall conductivity signal resulting from such state will be quantized; to see why this is in fact true, let us consider this HF many body state as defining an effective \textit{filled} band. For a spin-valley-band polarized state which is achieved at large interaction strength, the filled band coincides with one of the HWF bands and thus has manifestly a nonzero Chern number;
we can then consider a HF solution for smaller interaction strength, where the single particle states belonging to the effective filled band at each Bloch momentum could be written as linear combinations of the two HWF bands within a single valley; 
since the Chern number of a band is a topological property, one expects it to be invariant under smooth deformations of the band; 
starting from a spin-valley-band polarized state and decreasing the interaction strength, we expect that as long as the HF gap introduced above is not closed, the Chern number is intact and QAHE is expected.

A plot of such gaps as functions of interaction strength for several parameter choices is shown in Fig.~\ref{fig:gaps_eta_0}.
Note that the polarized state continues to exhibit HF stability as the interaction is lowered but becomes unstable when the interaction energy per particle becomes roughly comparable to the band width. 
Moreover, we consider a range of $\Delta$ from small to large values (always smaller than the noninteracting gap to remote bands); as shown in Fig.~\ref{fig:gaps_eta_0}, regardless of the value of $\Delta$, large interaction strength stabilizes the QAHE, while in the range of small interaction strength, larger $\Delta$ results in a more stable polarization.
In addition, at intermediate interaction strength, a second gap between HF eigenvalues, apart from the one induced by the interaction, is visible due to the large sublattice potential and scales with $\Delta$ 
(see Fig.~\ref{fig:energies_eta_0_Delta_01}); obviously, we will keep track of the former to study stability of QAHE. Also, as is expected and also shown in Fig.~\ref{fig:gaps_eta_0}, tuning the twist angle away from the magic value results in weaker stabilization of QAHE and generally larger interaction is needed to stabilize the QAHE.
 
\begin{figure}[!t]
	\centering
	\subfigure[]{\label{fig:gaps_eta_0_changing_lxi}\includegraphics[height=0.225\textwidth]{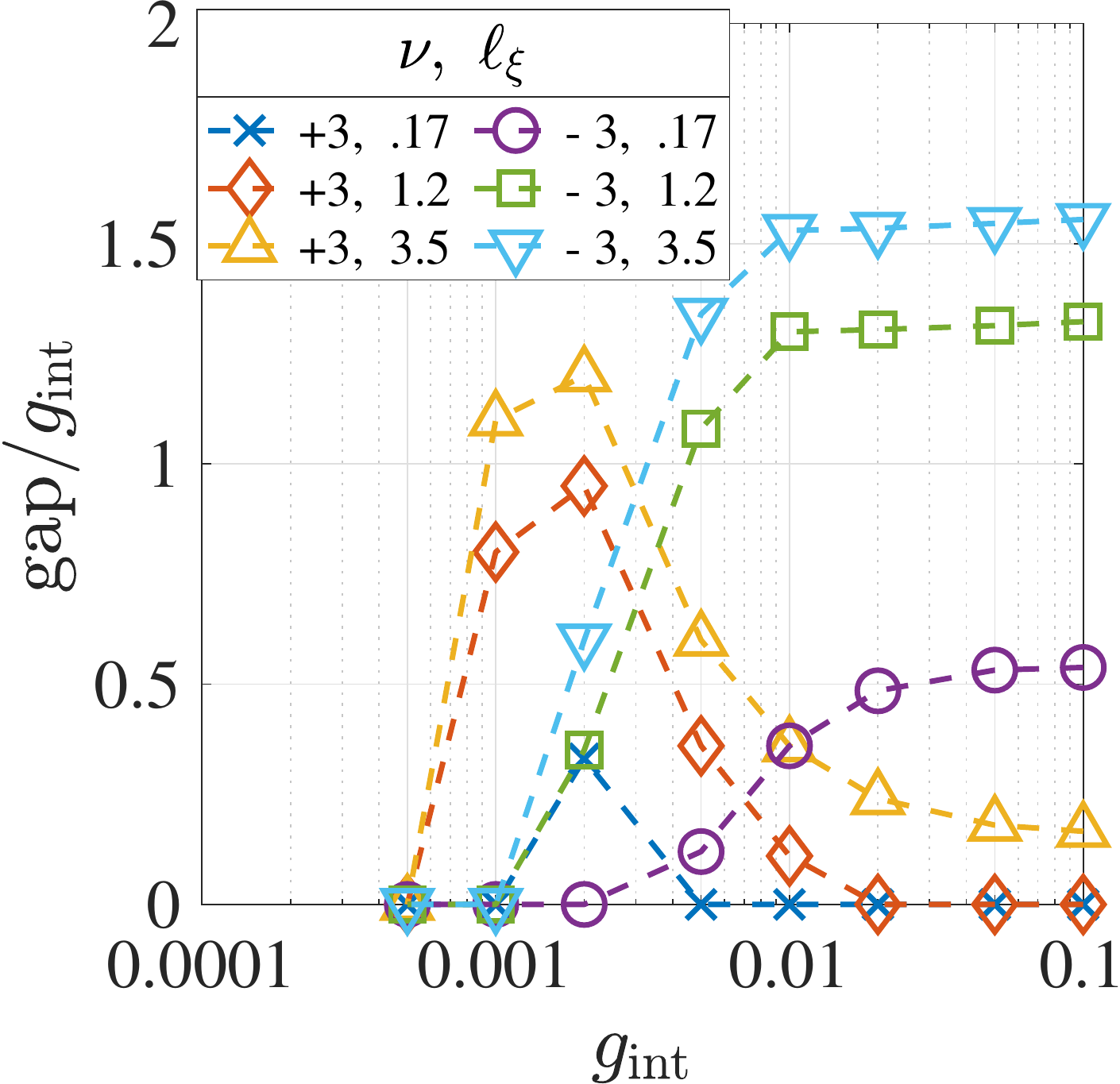}}
	\subfigure[]{\label{fig:gaps_eta_changing}\includegraphics[height=0.225\textwidth]{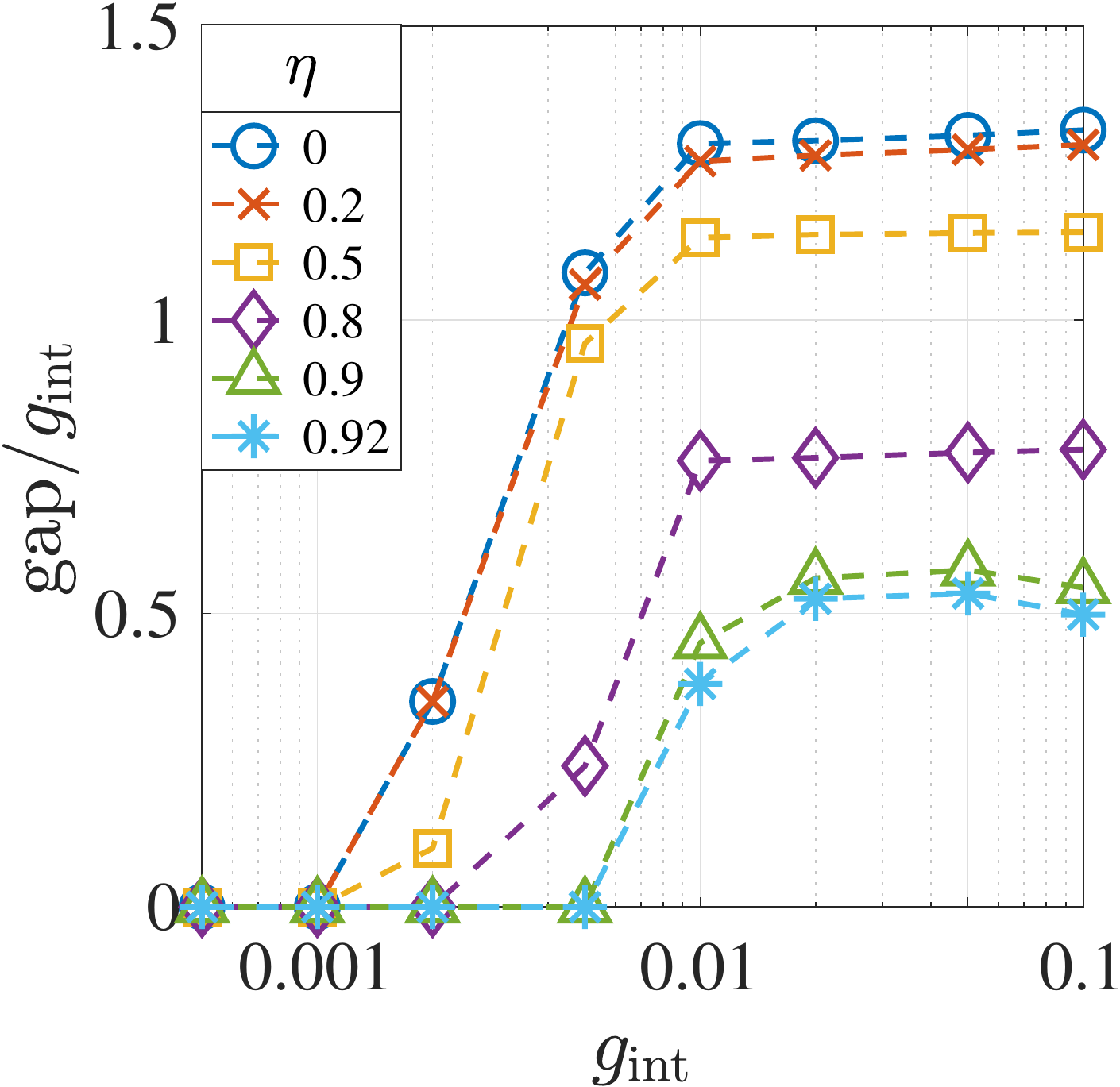}}
	\caption{\footnotesize Normalized gaps as functions of $g_{\text{int}}$ with longer range interaction and nonzero $\eta$ in the first study. $\alpha = 0.58$, and $\Delta = 1.9 \text{meV}$ have been chosen. (a) Here both of the fillings are considered still in the chiral limit ($\eta=0$). The QAHE is more stable at larger screening length; interestingly, it is stabilized even for $\nu=+3$ for large interaction, if large enough screening length is chosen. Note that this feature is lost for larger $\eta$ values and in particular $\eta_{\text{phys}} = 0.8$, as discussed in the main text. (b) Gaps are drawn for $\nu = -3$ with $\ell_\xi = 1.2 a_M$ here, as $\eta$ is increased. Increasing $\eta$ makes QAHE less stable until it is not stabilized at all even at large $g_{\text{int}}$ at $\eta \approx 0.9 - 0.95$. 
	}
	\label{fig:eta_0_lxi_and_eta_nonzero}
\end{figure}

 At the filling of $+3$, on the other hand, starting from large values of interaction strength, with the present settings, the QAHE state is not stabilized. However, upon decreasing the interaction strength, interestingly, when the interaction energy per particle becomes comparable to the band width, a narrow interval of interaction strength allows for the QAHE to be stabilized although it gets unstable again for smaller interactions (see Fig.~\ref{fig:gaps_eta_0_changing_lxi}). This observation holds true irrespective of the value of $\Delta$.

The above discrepancy between the two filling factors indicates that there is a particle-hole {\it asymmetry} in the system with the current choice of the Hamiltonian, although the non-interacting Hamiltonian is chiral and thus particle-hole symmetric with and without $\Delta$. This asymmetry could be understood by noting the following fact within the active-bands-only model we have chosen to work with here, i.e.~the choice of $H_{\text{MF},0} = 0$: 
starting from the extreme cases, there is a difference between a single electron at $\nu = -4$ and a single hole at $\nu = +4$, in that, the hole senses an additional potential due to the presence of eight full bands of electrons.
In the same fashion, a single hole senses an additional $k$-dependent potential at $\nu = +3$ when compared with an electron at $\nu = -3$, and thus some $k$ values in the hole bands could be preferred over others; more details can be found in Appendix \ref{app:ph_hwf}. This single hole potential is interaction induced and thus becomes stronger as the interaction is raised. One can argue that destabilization of QAHE in the strong interaction limit of the filling $+3$ presented above happens exactly due to this potential; holes prefer to occupy some momenta more than others. As we will see below, using a longer range interaction could weaken this asymmetry.

Let us mention two important points regarding the particle hole transformation of the many body state here before moving on: our choice of $H_{\text{MF},0} = 0$ here means that single electron excitations on top of $\nu = -4$, receive no HF correction in their dispersion. Had we chosen another form for $H_{\text{MF},0}$, so that the holes at $\nu = +4$ experienced no change in dispersion from the CM, we would have gotten the same theory but with electrons replaced with holes; in other words, using this prescription for $H_{\text{MF},0}$ will yield the particle hole transformed version of the present model with $H_{\text{MF},0} = 0$. Additionally here we only discussed the model at $\eta = 0$, where there is a chiral symmetry in the model, while for generic $\eta$, there is an approximate particle hole symmetry in the CM which plays a similar role. With this particle hole symmetry one can repeat the above considerations for nonzero $\eta$ as well, i.e.~show that the symmetry between holes and electron at the two fillings $\pm 3$ is broken within the present model and also that in a particle hole transformed version of the model, holes will play the role of electrons (see Appendix \ref{app:ph_hwf} for details).

\subsubsection{$\eta = 0$, longer range interaction}
Upon using longer range interactions, which are more realistic, some of the results presented above are altered: a longer range interaction does not change the picture at $\nu=-3$ much, i.e.~with the use of longer range interactions, the QAHE is still stabilized at large interaction and stability is lost at small enough interaction strength (see Fig.~\ref{fig:gaps_eta_0_changing_lxi}). However at the filling $+3$ the effect is more remarkable: the narrow range of the polarized states is made wider. As can be seen in Fig.~\ref{fig:gaps_eta_0_changing_lxi}, above some intermediate screening length, even at large interaction the QAHE state is stabilized. Generally, we have observed that increasing $\ell_\xi$ makes QAHE more stable.

\subsubsection{$\eta \neq 0$, away from the chiral limit}
We take another step toward making the model more realistic by choosing $\eta$ to be nonzero and increasing it to the physical value $\eta_{\text{phys}} \approx 0.8$\cite{koshino2018maximally}; the physics at $\nu = -3$ stays similar to a high extent even up to $\eta_{\text{phys}}$. However, at larger $\eta$, i.e. $\eta \approx 0.9-0.95$, one starts to observe that the HF iterations do not stabilize the QAHE at this filling even with largest interactions. This means that the spin-valley polarized state ceases to be a local minimum in energy (among Slater determinant states) even when the interaction plays the dominant role. As can be observed in Fig.~\ref{fig:gaps_eta_changing} the HF gap becomes smaller as $\eta$ is increased. For the filling of $+3$, on the other hand, we observed that although large interaction of long enough range stabilizes QAHE at small $\eta$, for larger $\eta$ and in particular for the physical value this ceases to be true no matter how long range the interaction is chosen.

\subsubsection{symmetry transformed states}
Before closing the discussion of our first study, we would like to comment on other QAHE states that are obtained by symmetry transformations on the nearly spin-valley-band polarized ones. We start by considering the chiral limit, as was mentioned earlier a U(4)$\times$U(4) symmetry of separate transformations of the two Chern sectors is seen in the interaction part of the Hamiltonian\cite{bultinck2019ground}; as is discussed in Appendix \ref{app:hamiltonian_hwf}, when the kinetic terms are also considered in the chiral limit, the symmetry of the total Hamiltonian reduces to U(4); this is due to the fact that the unitaries acting in the two Chern sectors cannot be chosen independently (see Appendix \ref{app:hamiltonian_hwf}). Apart from nearly polarized states in our numerics, we also observe states obtained by acting with such transformions on the nearly polarized states. The above intra-Chern-sector symmetry does not survive moving away from the chiral limit. It is also worthwhile to mention that apart from the ones discussed above, we did not obtain any other solution to our unrestricted HF calculations.

\bigskip

Next, we turn our attention to a second study with a projected Hamiltonian.

\subsection{Second study}
\label{sec:ss}

\begin{figure}[!t]
	\centering
	\subfigure[]{\label{fig:ph_hf_spectra}\includegraphics[height=0.29\textwidth]{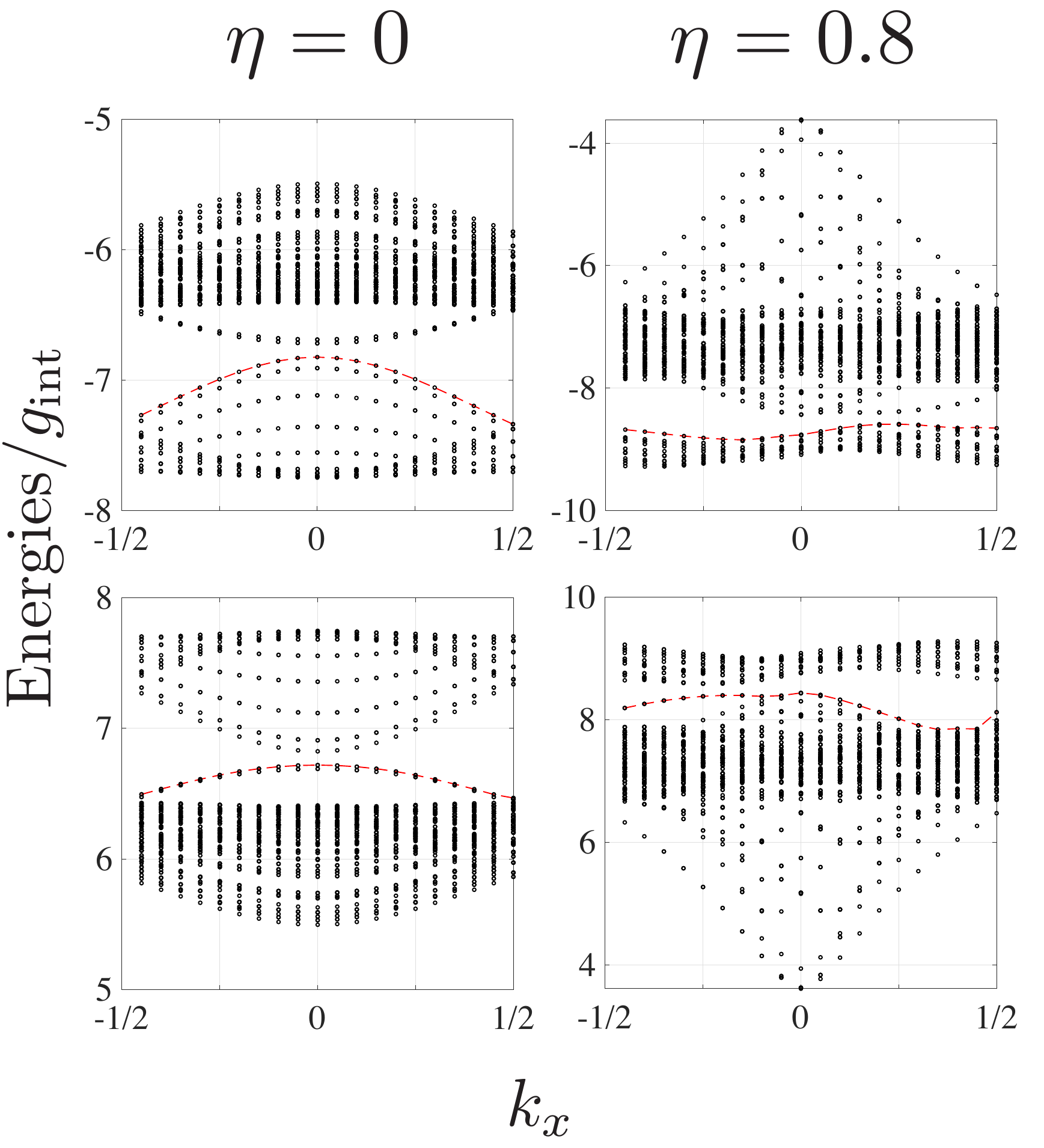}}
	\subfigure[]{\label{fig:gaps_second_study}\includegraphics[height=0.225\textwidth]{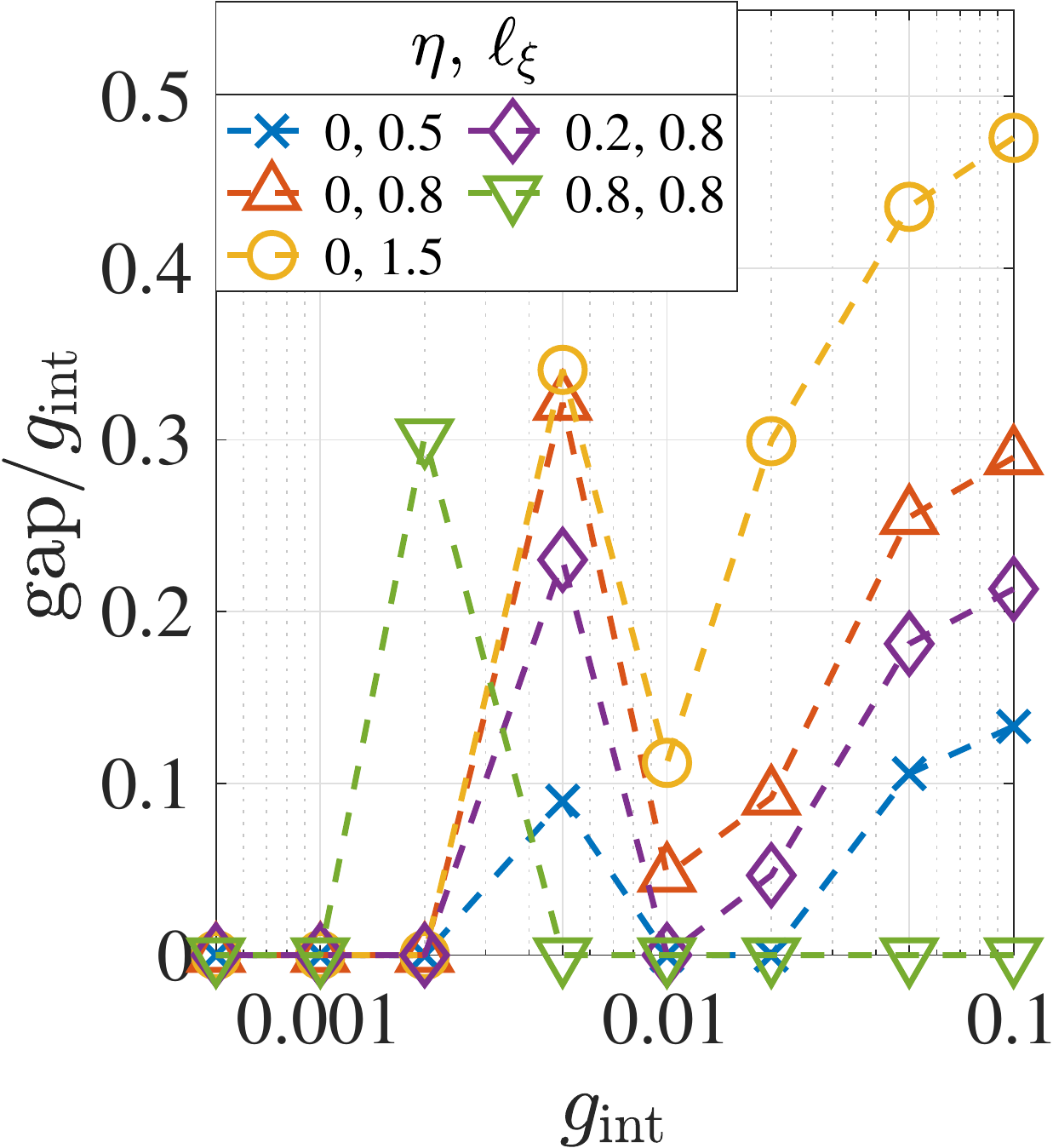}}	
	\caption{\footnotesize The HF results obtained within the projected model, i.e.~the model used in our second study. (a) The converged HF energies normalized by $g_{\text{int}}$  vs $k_x$. In all these four subplots, we take $\alpha = 0.58, \Delta = 0, \ell_\xi = 0.5 a_M$. The interaction strength is chosen as $g_{\text{int}} = 0.05$ and $0.002$ for the left and right columns respectively. Plots in each column have the same parameter values except for the filling factor: the first row corresponds to $\nu = -3$ and the second row to $\nu = +3$. It can be seen that the HF energies at the two fillings $\nu = \pm3$ can be transformed into each other by the particle hole symmetry of the CM, which in particular needs $k_x \to - k_x$.
	It can be seen that for $\eta = 0$, the two sets of HF energies are also related by the chiral symmetry of the CM.  
	(b) The gaps as functions of $g_{\text{int}}$. $\alpha = 0.58, \Delta =0$ are chosen for this plot and $\eta$ and $\ell_\xi$ are varied. 
	}
	\label{fig:second_study_plots}
\end{figure}

In this subsection we work with a Hamiltonian that is obtained by projecting an interacting Hamiltonian onto the subspace of active bands, and the zero point of the HF approach is chosen to be at the CNP of the moir\'e bands, i.e.~we will use Eq.~\eqref{eq:H_MF_second_approach}. Unlike the previous case, this choice results in a particle hole symmetry between the many body states at the two filling factors $+ \nu$ and $-\nu$ (see Appendix \ref{app:ph_hwf} for details), and therefore we will focus on $\nu = -3$ only in this study. Note that this particle hole symmetry is present regardless of the value of $\eta$ and is reflected in the HF spectrum. As an illustration, we present two sets of converged HF energies shown in Fig.~\ref{fig:ph_hf_spectra} with $\eta = 0 $ and $\eta = 0.8$. The HF energies at $\nu=+3$ and $\nu=-3$ are related by the particle hole transformation which in particular involves a $k_x \to - k_x$ transformation. Notice that for $\eta=0$, these two sets of HF energies are also related by the chiral symmetry.

Fig.~\ref{fig:gaps_second_study} shows the HF gap as a function of interaction strength for several parameter choices.
We observe that for large interaction strength at small $\eta$, QAHE is stabilized under HF iterations but this does not happen for larger $\eta$.  In particular, the QAHE phase is HF stable at $\eta_{\text{phys}} = 0.8$ for a small windows of parameter choices and is absent when $\eta=0.5$.
We generically see a bump in the rescaled gap as shown in Fig.~\ref{fig:gaps_second_study} when interaction strength is comparable to the noninteracting energies. This occurs due to a partial cancellation between quadratic terms of the Hamiltonian that arise from $H_{\text{kin}}$ and $H_{\text{MF},0}$; it is indeed this same effect that gives rise to the narrow window exhibiting QAHE at $\eta_{\text{phys}}$. Furthermore, the destabilization of the QAHE at larger $\eta$ values
for large $g_{\text{int}}$ is attributed to the fact that $H_{\text{MF},0}$ has quadratic terms that scale with $g_{\text{int}}$, and these terms prefer states with particular $k_x$ values over others.

The above results have focused on the $\Delta = 0$ limit. One can also consider finite $\Delta$ or even the large $\Delta$ limit. The latter limit is defined by requiring that the noninteracting gap due to $\Delta$ is not closed by the interaction induced effects,
hence schematically $g_{\text{int}} \lessapprox \Delta$. But this limit also results in similar behavior of the HF stability to the present model and we will not present these numerical results here.

\subsection*{Discussion}
We have considered two different models with different $H_{\text{MF},0}$, in a manner that the HF zero point is chosen at $\nu=-4$ and $\nu=0$ (CNP) respectively. The former case, as is discussed in Appendix \ref{app:ph_hwf}, is actually also related to a model with HF zero point choice of $\nu=+4$, by a particle hole transformation. More specifically, the many body states at the filling factor $\nu$ that are obtained using the HF zero point of $\nu=-4$, with an appropriate replacement of electrons with holes, are equivalent to  states at filling factor $-\nu$, if the HF zero point is moved to $\nu = +4$.  On the contrary, the choice of $\nu = 0$ as the zero point, i.e.~the case in the second study, is always particle-hole symmetric.

We would like to emphasize that the particle hole symmetry discussed above is expected to be broken even at the noninteracting level when actual physical effects like lattice relaxation are taken into account. However, we should bear in mind that if the particle hole symmetry is not broken at the noninteracting level, the interactions will also keep it intact.
On the other hand, the current experiments exhibiting QAHE\cite{serlin2020intrinsic,Sharpe605} only observe the effect at $\nu=+3$, and not at $\nu=-3$, which is an indication of particle hole symmetry breaking.
On a phenomenological level, this makes us speculate  that among the three different cases discussed above, the HF zero point choice of $\nu = +4$ is probably most relevant to the physics seen in the samples exhibiting QAHE.
Let us mention that ultimately within the framework of this paper, we cannot argue in favor of any of the above three choices. However we note that a definitive answer to this issue needs further study of several other effects that are neglected here; in particular, consideration of the particle hole symmetry breaking effects (such as lattice relaxation as mentioned above) and also more careful treatment of the effects of the filled remote bands could play decisive roles in determining which of the above choices (if any) could serve as a consistent physical model describing the relevant physics.

\section{Conclusion}
\label{sec:conclusion}

To summarize, we introduce the hybrid Wannier basis in the continuum model of TBG and study the strong interaction effect by using the self consistent Hartree Fock approximation. We focus on the filling factors $\pm 3$ and investigate the stability of QAHE phases at these two fillings. 
Interestingly, we observe that stability of the QAHE depends crucially on the zero point choice of the HF dispersion. In the range of physically relevant parameter choices we see that the QAHE is most robustly stabilized at large interaction strength under HF for the zero point choices of $\pm 4$, and the corresponding filling factors of $\nu = \pm 3$. 
We note that the QAHE is observed in experiments on magic angle TBG at $\nu = +3$.
Moreover, we numerically observe that reducing the sublattice potential, reducing the screening length and tuning away from the magic angle, generically make the QAHE less stable.  In particular, the weakened stability by reducing the sublattice potential is consistent with experiments \cite{serlin2020intrinsic,Sharpe605}, which have observed the effect in TBG samples with aligned hexagonal Boron Nitride substrates, which is believed to induce a sublattice potential.

Further development of the present method can be envisioned.  The HWF basis we introduced in this paper might be used to find possible fractional phases at non-integer fillings and other interesting phases at integer filling factors. Another possible application, for which HWFs are particularly well suited, is to address situations containing spatially varying configurations such as domain walls between different symmetry broken states or states with one-dimensional ``stripey" translational symmetry breaking. We leave these directions for the future study.

\begin{acknowledgements}  
KH thanks O. Vafek, J. Kang, A.F. Young, M. Serlin, C. Repellin, C. Liu, C. Tschirhart, F. Schindler for fruitful discussions, and in particular appreciates numerous helpful conversations with Y.~Alavirad. 
We acknowledge support from the Center for Scientific Computing from the CNSI, MRL: an NSF MRSEC (DMR-1720256) and NSF CNS-1725797. The research of KH and LB was supported by the NSF CMMT program under Grants No. DMR-1818533, and we benefited from the facilities of the KITP, NSF grant PHY-1748958. XC acknowledges support from the DARPA DRINQS program.

\end{acknowledgements}

\bibliography{graphene}

\onecolumngrid
\appendix

\section{The model}\label{app:model}
In this Appendix, we briefly review the continuum model of Bistritzer and MacDonald, which is the starting point of this work. We take the Hamiltonian for the two valleys ($\xi=\pm 1$) as
	\begin{equation}\label{eq:TBG_no_magnetic}
	\begin{aligned}
		H(\bm{x}) &= -i\, \left(\bm{\nabla} + i\xi \left[ - \tau^z \frac{\bm{q}_0}{2} + \bm{q}_{\text{h}} \right] \right) \cdot \left( \xi \sigma^x , \sigma^y \right) \\
		&+ \alpha \; \tau^+ \left[\eta \, \beta_0(\bm{x})+\beta_1(\bm{x}) \sigma^+ + \beta_2(\bm{x}) \sigma^-\right] +\mathrm{h.c.}.
	\end{aligned}
\end{equation}
The Hamiltonian and the position are made dimensionless by dividing by $\hbar v_F k_\theta$ and $\frac{1}{k_\theta} = \frac{3a_M}{4\pi }$, where $a_M = \frac{\sqrt{3}a}{\theta} $ is the moir\'e unit length ($a$ is the distance between adjacent carbon atoms in graphene). The Pauli matrices $\tau^z, \sigma^z$ denote the layer and sublattice degrees of freedom. There are two parameters in the above Hamiltonian: $\alpha = \frac{w_{\text{AB}}}{\hbar v_F k_\theta} \sim \frac{w_{\text{AB}}}{v_F \theta}$ and $\eta = \frac{w_{\text{AA}}}{w_{\text{AB}}}$. The moir\'e periodic funcitons $\beta_n(\bm{x}) = \sum_{j=0}^2 e^{-i \left(\xi\bm{Q}_j\right) \cdot \bm{x}} \zeta^{\xi nj}$, are defined in terms of moir\'e reciprocal lattice vectors $\bm{Q}_0 = 0, \quad \bm{Q}_1 = \sqrt{3} \left( -\frac12,\frac{\sqrt{3}}{2} \right), \bm{Q}_2 = \sqrt{3} \left( \frac12,\frac{\sqrt{3}}{2} \right)$, and also with $\zeta = e^{2\pi i/3}$.
Also, $\bm{q}_{\text{h}}=\left(\frac{\sqrt{3}}{2},0\right)$, and $\bm{q}_0 = \left( 0 , -1 \right)$.  Notice that we have neglected the opposite rotation of sublattice matrices in the two layers in the Hamiltonian \eqref{eq:TBG_no_magnetic}, which is a small effect (order $\theta$) for small twist angles; this results in an approximate particle/hole symmetry which is detailed below along with some other symmetries of the CM. 
 
The presence of a $C_2\mathcal{T}$ breaking sublattice potential is also considered in this work which is taken to be of the form $\Delta \, \sigma^z$.
 
 \vspace{0.5cm}
 
 \textbf{Some symmetries of the CM:}
\begin{itemize}
    \item The neglect of the rotation of the sublattice matrices results in a particle-hole symmetry as defined in Ref.~\onlinecite{hejazi2019multiple} (see also Ref.~\onlinecite{song2019all}), note that it is an intravalley transformation:
\begin{equation}
U^\dag_{\text{ph}} H_{(-k_x,k_y)}\Big((-x,y)\Big)U^{\phantom\dag}_{\text{ph}} = -H_{\bm{k}}(\bm{x}),
\end{equation}
where $U_{\text{ph}} = \sigma^x \tau^z e^{2i \left(\xi \bm{q}_{\text{h}}\right)\cdot\bm{x}}$. This symmetry is preserved even if a sublattice symmetry breaking term is also present. One should have in mind that this particle-hole symmetry is different from the chiral symmetry of the Chiral model, since the latter keeps the Bloch momentum intact and the present particle-hole symmetry takes $(k_x,k_y) \to (-k_x,k_y)$ and also is present regardless of the value of $\eta$ within the above approximation. 

\item  The above form of the Hamiltonian (no sublattice potential) has a $C_2\mathcal{T}$ symmetry, which also acts within a single valley:
\begin{equation}
U^\dag_{C_2\mathcal{T}} H^*_{\bm{k}}(-\bm{x})U^{\phantom\dag}_{C_2\mathcal{T}} = H_{\bm{k}}(\bm{x}),
\end{equation}
with $U_{C_2\mathcal{T}} = \sigma^x$. In the plane waves basis, it has the form $\sigma^x \mathcal{K}$, where $\mathcal{K}$ is the complex conjugation.

\item There is another intravalley symmetry of our interest, which is a mirror symmetry with respect to $y$:
\begin{equation}
    U^\dag_{M_y} H_{( k_x , -k_y )}\Big( ( x , -y ) \Big)U^{\phantom\dag}_{M_y} = H_{\bm{k}}(\bm{x}),
\end{equation}
where $U_{M_y} = \sigma^x \tau^x$. This symmetry is also only present when $\Delta = 0$, and thus when $C_2\mathcal{T}$ is not broken.

\item There is a time reversal symmetry that acts between the two valleys:
\begin{equation}
     H_{-\bm{k}}(\bm{x}) \Big|_{\xi = -1}^* = H_{\bm{k}}(\bm{x}) \Big|_{\xi = +1} 
\end{equation}

\end{itemize}

\section{Maximally localized hybrid Wannier functions}\label{app:max_local_HWF}
As discussed in the main text the maximally localized HWFs are actually maximally localized one dimensional Wannier functions for each $k_x$ that are derived using the method in Ref.~\onlinecite{marzari1997maximally}. In this Appendix a sketch of the procedure is presented, and also special cases are discussed in more length.

At each $k_x$, the spread function 
$$\Omega_{k_x,\xi} = \sum_m \left[ \left\langle y^2  \right\rangle_{m,k_x,\xi} - \langle y  \rangle_{m,k_x,\xi}^2 \right],$$
where the expectation values are calculated with respect to states
$ \left| k_x ; y_c , m ,\xi  \rangle \right.$, is minimized through suitable gauge transformations of the Bloch functions; the spread function consists of an invariant part $\Omega_{\text{I}}$, which does not change under gauge transformation at all, and a contribution which can be minimized; the latter on its own comprises a band-diagonal part $\Omega_{\text{D}} = \frac1{N_y}\sum_{k_y} \frac1{b^2}\sum_m \left(-\mathrm{Im} \log \mathcal{M}^{\bm{k},\xi}_{mm} - b_y \langle y \rangle_{m,k_x,\xi} \right)^2$ and a band-off-diagonal part $\Omega_{\text{OD}} = \frac1{N_y}\sum_{k_y} \frac1{b^2}\sum_{m\neq m'} \left| \mathcal{M}_{mm'}^{\bm{k},\xi}  \right|^2$. Here we give more details for the procedure discussed in the main text; starting from a smooth gauge for the original Bloch functions, suitable off-diagonal gauge transformations are made so that the $\mathcal{M}$ matrices are updated to be Hermitian. This is done by making use of the singular value decompositions (SVD) of the $\mathcal{M}$ matrices as follows (for every $\mathcal{M}$, one can define the SVD to have the form $\mathcal{M} = V \Sigma W^\dagger$, where $V$ and $W$ are unitary and $\Sigma$ is diagonal with nonnegative entries): starting from a point in the BZ for every $k_x$, say $k_{y,0} = -\frac32$ or in other words the left edge of the rectangular BZ, one can do series of gauge transformations separately along each constant $k_x$ line, so that all (except for the last one completing the 1D loop) $\mathcal{M}$ matrices become Hermitian; this is done by the Gauge transformation 
\begin{equation}
    \left( \left| u_{k_x,k_y;1,\xi} \right\rangle , \left| u_{k_x,k_y;2,\xi} \right\rangle \right) \to \left( \left| u_{k_x,k_y;1,\xi} \right\rangle , \left| u_{k_x,k_y;2,\xi} \right\rangle \right) \cdot 
    \left[\left( W V^\dagger   \right)_{k_y-by} \ldots \left( W V^\dagger   \right)_{k_{y,0}} \right],
\end{equation}
where $\cdot$ denotes a matrix multiplication in the space of bands, and the $k_x,\xi$ indices on $W$ and $V$ matrices are suppressed. In traversing the BZ in the $y$ direction once, one is able to define the accumulated matrix
\begin{equation}
    \Lambda_{k_x,\xi} = \left[\left( W V^\dagger   \right)_{-k_{y,0}-by} \ldots \left( W V^\dagger   \right)_{k_{y,0}} \right],
\end{equation}
note that this matrix naively gives the prescription for a change of basis at $k_{y,0}$, the point one started with. However, we would like to end up with the state we started with so that a smooth Bloch basis is defined throughout the 1D Brillouin zone. One can achieve this if a series of actions are taken: at all $k_y$ points, a unique basis change is made with the unitary matrix that diagonalizes $\Lambda_{k_x,\xi}$, i.e.~the matrix $V_\lambda$, where $V_\lambda^\dagger \Lambda V_\lambda = \lambda$ with $\lambda$ a diagonal matrix.

This last basis change updates all of the $\mathcal{M}$ matrices (except the last one at $-k_{y,0}-b_y$, more on this below) to have the form of a Hermitian matrix.
The Hermitian matrices are proportional to unity to first order in $b_y$, and this ensures that $\Omega_{\text{OD}}$ shown above vanishes to first order in lattice spacings. However, there remains band-diagonal total Berry phases in this new basis which are invariant under single band gauge transformations; these are the inverses of the eigenvalues of the $\Lambda$ matrix defined above and are at this stage accumulated in the last $\mathcal{M}$ matrix, i.e.~at $-k_{y,0}-b_y$. One should make a band-diagonal gauge transformation (a phase redefinition) to ensure that this Berry phase is distributed evenly along the one-dimensional Brillouin Zone to make $\Omega_{\text{D}}$ vanish as well. This final (band diagonal) gauge transformation results in the final form $K\gamma$ for the $\mathcal{M}$ matrices, with a Hermitian $K$ and a diagonal unitary $\gamma$ for the $\mathcal{M}$ matrices.

Note that an evenly distributed Berry phase means that $\gamma^{k_x,\xi}$ is independent of $k_y$ and in fact equal to $\lambda^{-\frac1{N_y}}$. Furthermore, bear in mind that the matrix $\Lambda^\dagger$ is equal to the path ordered product of $\mathcal{M}$ matrices to first order in $b_y$ for each $k_x$ and thus is equal to the Wilson loop at $k_x$ to this order. Noting that eigenvalues of the Wilson loop operators are related to the WCCs of the final bands means that eigenvalues of $\gamma^{k_x,\xi}$ take the form $e^{\frac{2\pi i}{N_y} \; \overline{y_{k_x,n}}}$, where $\overline{y_{k_x,n}}$ denotes the Wannier charge centers at $k_x$ in units of $\frac12 a_M$.

The $U$ matrices defined in Eq.~\eqref{eq:bloch_redefinition}, can be explicitly derived as:
\begin{equation}
    U^{k_x,k_y;\xi} = \left[\left( W V^\dagger   \right)_{k_y-by} \ldots \left( W V^\dagger   \right)_{k_{y,0}} \right] \cdot V_\lambda \cdot \left( \lambda \right)^{-\frac{k_y - k_{y,0}}{2 k_{y,0}}},
\end{equation}
where all right hand side matrices are evaluated at $k_x, \xi$.

	\vspace{0.5cm}
	
Finally, we discuss further the special cases mentioned in the main text:
\begin{itemize}
    \item In the case where $\Delta = 0$, due to the $C_2\mathcal{T}$ symmetry of the Hamiltonian, one can work with Bloch eigenstates of Hamiltonian that are $C_2\mathcal{T}$ symmetric. Any inner product of two $C_2\mathcal{T}$ eigenstates is real; this means that the $\mathcal{M}$ matrices have the form $\exp\left[ i\mu^y m_{\bm{k},\xi} b_y\right] + \mathcal{O}(b_y^2)$, where $\mu^y$ acts in the two dimensional band space. Thus every SVD operator $VW^\dagger$ could be taken to be equal to $\mathcal{M}=\exp\left[ i\mu^y m_{\bm{k},\xi} b_y\right] + \mathcal{O}(b_y^2)$ and furthermore $V_\lambda$ could be taken to be the matrix that diagonalizes $\mu^y$. Additionally, the integrals $\pm \int dk_y \; m_{\bm{k},\xi}$ yield the single band total Berry phases of the two bands in the parallel transport basis which should be distributed evenly along the strip with $k_x$. All this means that the states $\frac{e^{ \pm i \phi_{\bm{k},\xi}}}{\sqrt{2}}   \left( \left| \psi_{\bm{k};1,\xi} \rangle  \right. \pm i \left| \psi_{\bm{k};2,\xi} \rangle \right. \right) $, with $C_2 \mathcal{T} \left| \psi_{\bm{k};m,\xi} \right\rangle$ = $\left| \psi_{\bm{k};m,\xi} \right\rangle$, form the parallel transport basis, if the phases are chosen properly to distribute the single band Berry phases evenly along the $y$ direction. 
    \item In the case of $\eta = 0$, regardless of the value of $\Delta$, the sublattice polarized states form the parallel transport basis. One can argue for this as follows: starting from $\Delta = 0$, we note that states with opposite sublattice polarizations automatically have zero contribution to $\Omega_{\text{OD}}$. Suitable single band gauge transformations are furthermore needed to minimize $\Omega_{\text{D}}$ as well. On the other hand, we know that the two bands in the chiral limit are related by\cite{tarnopolsky2019origin}: 
	$\left| \psi_{\bm{k} , \xi , 1} \right. \rangle = i \, \sigma^z \left| \psi_{\bm{k} , \xi , 2} \right. \rangle.$
    This means that adding the term $\sigma^z\Delta$ to the Hamiltonian does not change the subspace of active bands. And thus previously found sublattice polarized states still represent the active bands subspace, and with suitable single band phase redefinitions will form the parallel transport basis. It is important to note that addition of $\Delta$ does not change Wilson loop matrices for each $k_x$, and thus the phases chosen for $\Delta = 0$ in the parallel transport basis remain valid choices for nonzero $\Delta$ as well.
\end{itemize}

\section{Hamiltonian in the HWF basis} \label{app:hamiltonian_hwf}
In this section we describe how different terms of the Hamiltonian are derived in the HWF basis.

\begin{itemize}
    \item \textbf{Kinetic term:}
\end{itemize}
$H_{\text{kin}}$ could be written in different bases, we start by writing it in the basis of original Bloch eigenstates:
	\begin{equation}
	\begin{aligned}
		E^{\bm{k};\xi} &= 
		\frac{1}{N_x N_y}\begin{pmatrix}
			\left\langle  \psi_{\bm{k} ; 1 , \xi} \right|  \\
			\left\langle  \psi_{\bm{k} ; 2 , \xi} \right|
		\end{pmatrix}
		\ H_{\text{kin}} \
		\left(
			\left|  \psi_{\bm{k} ; 1 , \xi} \right \rangle ,
			\left|  \psi_{\bm{k} ; 2 , \xi} \right \rangle 
		\right) \\
		&= \begin{pmatrix}
			E^{\bm{k};\xi}_1 & 0\\
			0 & E^{\bm{k};\xi}_2
		\end{pmatrix}.
	\end{aligned}
	\end{equation}
This defines the diagonal energy matrix $E^{\bm{k};\xi} $. The kinetic term in the HWF basis then reads:
	\begin{equation}
	\begin{aligned}
		t^{y_c'-y_c \, ; \, k_x' , k_x \, ; \, \xi' , \xi} &= \frac{1}{N_x}  \begin{pmatrix}
			\left\langle  k_x' ; y_c' , 1 , \xi'\right|  \\
			\left\langle  k_x' ; y_c' , 2 , \xi' \right|
		\end{pmatrix}
		\ H_{\text{kin}} \
		\left(
			\left|  k_x ; y_c , 1 , \xi \right \rangle ,
			\left|  k_x ; y_c , 2 , \xi \right \rangle  
		\right)   \\
		&= \delta_{k_x' \, k_x} \delta_{\xi' \, \xi} \left\{ \frac{1}{N_y}  \sum_{k_y} e^{ik_y(y_c'-y_c)}   \left[ \left( U^{\bm{k},\xi} \right)^\dagger \ E^{k_x,k_y;\xi} \ U^{\bm{k},\xi}  \right]\right\}
	\end{aligned}
	\end{equation}
	and this defines the hopping matrix. As a result the kinetic term can be written as:
	\begin{equation}
		H_{\text{kin}} = \sum_{k_x,y_c',y_c,m',m,\xi,s} \left| k_x ; y_c' , m' , \xi , s\right \rangle    \left\langle  k_x ; y_c , m , \xi , s \right|  \ \ t^{y_c'-y_c \, ; \, k_x  \, ; \, \xi}_{m'm}
	\end{equation}
	Where spin index has also been added trivially.

\vspace{0.5cm}

\begin{itemize}
    \item \textbf{Interaction terms:}
\end{itemize}

The electron-electron interactions involve all electrons regardless of which moir\'e bands of the CM they belong to. However, here we are making an assumption that the gap between the active bands and the remote bands is large compared to the electron-electron interactions and thus it is legitimate to take the active bands as rigidly empty or full.

First, we discuss the four Fermi interaction Hamiltonian between the electrons in the active bands in the HWF basis; it takes the following form, the notation will be changed from $y_c$ to $y$ in HWF indices:
	\begin{equation}\label{eq:interaction_explicit_HWF}
	\begin{aligned}
		H_{\text{int}} &= \frac12 \frac{1}{N_x^2} \sum_{[k_x] \, ; \, [y] \, ; \, [m]} \ \sum_{\xi,\xi',s,s'} \ \mathcal{I}_{[k_x] \, ; \, [y] \, ; \, [m] , \xi,\xi'} \ \\
		&\qquad \qquad c^{\dagger}_{k_{x,1}, y_1, m_1, \xi, s}\, c^{\dagger}_{k_{x,2}, y_2, m_2, \xi', s'} \, c^{\phantom{\dagger}}_{k_{x,3}, y_3, m_3, \xi', s'} \, c^{\phantom{\dagger}}_{k_{x,4}, y_4, m_4, \xi, s},
	\end{aligned} 
	\end{equation}
	with the coefficients shown by $\mathcal{I}$ as follows:
	\begin{equation}
	\begin{aligned}
		\mathcal{I}_{[k_x] \, ; \, [y] \, ; \, [m] , \xi,\xi'}  &= \frac{1}{N_y^2} \sum_{[k_y]} e^{i \left( k_{y,1} y_1 + k_{y,2} y_2 - k_{y,3} y_3 - k_{y,4} y_4\right)} 
		\\
	 & 
	 \qquad \bigg\{ \frac{1}{N_x N_y \mathcal{A}} \sum_{\bm{G}} \delta_{\bm{k}_1 + \bm{k}_2 - \bm{k}_3 - \bm{k}_4,\bm{G}}  \ \ 
	 \times \\
		& \qquad  \qquad 
		\left[ \sum_{\Delta \bm{G}} \tilde{V}(\bm{k}_1 - \bm{k}_4 - \Delta\bm{G})  \quad \lambda_{m_1,m_4,\xi} \left(\bm{k}_1,\bm{k}_4,\Delta \bm{G} \right) \lambda^*_{m_3,m_2,\xi'} \left(\bm{k}_3,\bm{k}_2, \Delta\bm{G} - \bm{G} \right)\right] \bigg\}.
	\end{aligned}
	\end{equation}
In the above equation, we take the electron electron potential to have the form $\tilde{V}(\bm{q}) = \frac{e^2}{4\pi \epsilon } \frac{2\pi}{\sqrt{q^2 + \mu^2}}$. Furthermore, the form factors are defined in terms of certain inner products of parallel transport basis: 
\begin{equation}
		\lambda_{m_1,m_4,\xi} \left(\bm{k}_1,\bm{k}_4,\Delta \bm{G} \right) = \sum_{\bm{G}_1} \sum_{\sigma\tau} \tilde{\phi}_{\bm{k}_1,m_1,\xi}^*(\bm{G}_1,\sigma\tau) \, \tilde{\phi}_{\bm{k}_4,m_4,\xi}(\bm{G}_1 + \Delta \bm{G},\sigma\tau),
\end{equation}
where, the $\tilde{\phi}$'s are coefficients for expansions of parallel transport Bloch states in terms of plane wave states, i.e.~$\left|  \tilde{\psi}_{ \bm{k} ; m , \xi} \right \rangle = \sqrt{N_x N_y} \sum_{\bm{G},\sigma\tau} \tilde{\phi}_{\bm{k},m,\xi}(\bm{G},\sigma\tau) \left|\psi_{\bm{k} + \bm{G} ,\sigma\tau,\xi} \right\rangle$.

Second, we discuss the terms shown in the main text by $H_{\text{MF},0}$. Although remote bands are not treated as dynamical, a proper projection of the interacting Hamiltonian onto the active bands needs inclusion of an induced mean field potential due to the filled remote bands on the electrons in the active bands. This contribution will have the form:
\begin{equation}\label{eq:mean_field_effect_of_filled}
	H_{\text{MF, ind}} = \sum_{\alpha\alpha'}\left[\sum_{\beta \beta'} \left( V_{\alpha,\beta,\beta',\alpha'} - V_{\alpha,\beta,\alpha',\beta'} \right) \right] c^\dagger_\alpha c^{\phantom{\dagger}}_{\alpha'},
\end{equation}
where $\alpha,\alpha'$ run over active bands and $\beta,\beta'$ run over remote bands below CNP.

In addition to that, as discussed in the main text, we have taken the zero point of the HF to be at the CNP of the moir\'e bands. This means that at CNP, the single electron/hole dispersions as given by the CM should be unaltered under HF. In order for this to be true, we subtract the HF effect of the moir\'e CNP noninteracting state from the Hamiltonian. The addition of these two effects will result in the form given in Eq.~\eqref{eq:H_MF_second_approach} in the main text for $H_{\text{MF},0}$. 

There is a subtlety in the projection approach outlined above; with the above projected model at hand, we have considered changing the interaction strength in our study presented in the main text, this alters the coefficients of both $H_{\text{int}}$ and $H_{\text{MF},0}$ (a change in the dielectric constant, for example, could result in this effect). However, such a change will result in a different single electron/hole potential according to \eqref{eq:mean_field_effect_of_filled}; in particular, the single layer Fermi velocity $v_F$ and the interlayer tunneling parameters $w_{\text{AA}},w_{\text{AB}}$ will be renormalized, and other single particle terms will be induced or altered, these can include for example nonlinearities in the single layer dispersion in general, etc.~. A change in $v_F, w_{\text{AA}},w_{\text{AB}}$ parameters will tune the system away from the magic angle range. In this work, we have assumed that such change could be compensated by a change in the twist angle so that the system is tuned back to the new magic value for the twist angle as the interaction strength is altered. We have furthermore assumed that other induced effect (such as the monolayer nonlinear dispersion) could also be corrected by some means or are negligible and do not result in an appreciable effect. These assumptions allows us to also change the interaction strength in the terms correcting the zero point of our HF, and we will be left with the form in Eq.~\eqref{eq:H_MF_second_approach} with the interaction strength altered.

\begin{itemize}
    \item \textbf{Symmetries:}
    
    \begin{itemize}
        \item The $C_2 \mathcal{T}$, when present, acts on the parallel transport basis as stated in the main text, transforms one band to the other with $\bm{k}$ unchanged:
        \begin{equation}\label{eq:c2T_parallel_transport}
            \left\langle \bm{r} , \sigma \tau  \left| \tilde{\psi}_{\bm{k}, m , \xi } \right. \right\rangle  = \left\langle - \bm{r} , \bar{\sigma} \tau  \left| \tilde{\psi}_{\bm{k} , \bar{m} , \xi } \right. \right\rangle^*  .
        \end{equation}
        This in turn implies:
        \begin{equation}\label{eq:c2T_wave_functions}
            \left\langle \bm{r} , \sigma \tau  \left| k_x , y , m , \xi \right\rangle \right. = \left\langle - \bm{r} , \bar{\sigma} \tau  \left| k_x , -y , \bar{m} , \xi   \right. \right\rangle^*  = \left\langle \left( 2y \hat{\bm{y}} - \bm{r} \right) , \bar{\sigma} \tau  \left| k_x , y , \bar{m} , \xi \right. \right\rangle^* 
        \end{equation}
        Note that $y$ is an integer times $\frac{a_M}{2}$ and we have used the translational properties shown in Fig.~\ref{fig:real_space_lattice_BZ}.
        \item The particle hole symmetry exchanges the two bands of the HWFs basis, taking $k_x$ to $-k_x$. In the parallel transport basis the states can be related by this transformation as follows:
        \begin{equation}
            \left\langle \bm{r} , \sigma \tau  \left| \tilde{\psi}_{\bm{k}, m , \xi } \right. \right\rangle = (-1)^m \, i \ \left\{ e^{-i (2\xi\bm{q}_{\text{h}}) \cdot \bm{r}} \ (-1)^{1+\tau} \left\langle (-x , y) , \bar{\sigma} \tau  \left| \tilde{\psi}_{(-k_x,k_y), \bar{m} , \xi } \right. \right\rangle \right\} .
        \end{equation}
        
        The factor $(-1)^m i$ in the above equation can be derived in the $C_2 \mathcal{T}$ symmetric case explicitly; it furthermore can be maintained in the $\Delta \neq 0$ case as well by appropriate phase  redefinitions. The above property, furthermore, results in the symmetry of WCC positions under $k_x \leftrightarrow -k_x$, as seen in Fig.~\ref{fig:wilson_lines}. 
        
        \item The time reversal symmetry also relates the HWF states in the valleys in the following fashion:
        \begin{equation}\label{eq:valley_T_wave_functions}
            \left\langle \bm{r} , \sigma \tau  \left| k_x , y , m , \xi \right\rangle \right. = \left\langle \bm{r} , \sigma \tau  \left| - k_x , y , m , \bar{\xi} \right. \right\rangle^* .
        \end{equation}
        This symmetry can also be viewed in the parallel transport basis as:
        \begin{equation}\label{eq:valley_T_parallel_transport}
            \left\langle \bm{r} , \sigma \tau  \left| \tilde{\psi}_{\bm{k}, m , \xi } \right. \right\rangle = \left\langle \bm{r} , \sigma \tau  \left| \tilde{\psi}_{-\bm{k}, m , \bar{\xi} } \right. \right\rangle^* .
        \end{equation}
    \end{itemize}
    
    \item \textbf{Extra symmetry of the interaction term:}

    Interestingly, when $C_2 \mathcal{T}$ is present, the above equations show that under simultaneous action of the symmetries $C_2\mathcal{T}$, particle hole, time reversal and $M_y$ (not exhibited above for the parallel transport basis) on a parallel transport band, one obtains  the other band with the same Chern number, i.e.~one with the parallel transport band number and valley number swapped; explicitly, it is straightforward to show that the wave functions in these two bands satisfy the following relation (we will use the parallel transport basis for the following argument and not the HWF basis):
\begin{equation}\label{eq:valley_T_parallel_transport}
    \left\langle \bm{r} , \sigma \tau  \left| \tilde{\psi}_{\bm{k}, m , \xi } \right. \right\rangle = (-1)^\tau e^{-i \left(2 \xi \bm{q}_{\text{h}} i \right) \cdot \bm{r}} \left[ (-1)^{\bar{m}} \left\langle \bm{r} , \bar{\sigma} \bar{\tau}  \left| \tilde{\psi}_{\bm{k}, \bar{m} , \bar{\xi} } \right. \right\rangle \right] .
\end{equation}
This means that if one acts with this intra-Chern-sector transformation on one creation and one annihilation operator with the same spin and valley indices in the interaction terms, 
	\begin{equation}\label{eq:interaction_explicit_parallel}
		c^{\dagger}_{\bm{k}_{1}, m_1, \xi, s}\, c^{\dagger}_{\bm{k}_{2}, m_2, \xi', s'} \, c^{\phantom{\dagger}}_{\bm{k}_{3}, m_3, \xi', s'} \, c^{\phantom{\dagger}}_{\bm{k}_{4}, m_4, \xi, s},
	\end{equation}
the matrix element of the interaction remains unchanged; this implies the existence of a symmetry of the interaction term of the Hamiltonian, which we discuss for the chiral limit and also away from the chiral limit separately below:
\begin{itemize}
    \item $\eta = 0$, magic angle: in the chiral limit, since the parallel transport basis is sublattice polarized, the interaction in Eq.~\eqref{eq:interaction_explicit_parallel} becomes of density-density type in the band index as well as the spin and valley indices. This, along with the above observation of the invariance of interaction matrix elements, implies that the interaction terms could be grouped together so that only fermion bilinear terms 
    $\begin{pmatrix} c^{\dagger}_{\bm{k}_1,1,K} & c^{\dagger}_{\bm{k}_1,2,K'} \end{pmatrix} \begin{pmatrix} c^{\phantom{\dagger}}_{\bm{k}_4,1,K} \\ c^{\phantom{\dagger}}_{\bm{k}_4,2,K'} \end{pmatrix}$ 
    and 
    $\begin{pmatrix} c^{\dagger}_{\bm{k}_1,2,K} & c^{\dagger}_{\bm{k}_1,1,K'} \end{pmatrix} \begin{pmatrix} c^{\phantom{\dagger}}_{\bm{k}_4,2,K} \\ c^{\phantom{\dagger}}_{\bm{k}_4,1,K'} \end{pmatrix}$
    appear in the four Fermi terms of the interaction, where the spin indices are suppressed.
    This means that separate unitary transformations within each Chern sector keep the interaction intact. Upon further including the spin rotation symmetry as well, one recovers the two separate $U(4) \times U(4)$ symmetries of the two Chern sectors discussed in Ref.~\onlinecite{bultinck2019ground}.
    \item $\eta = 0$, away from the magic angle: in the chiral limit, the non-interacting Hamiltonian can be written in the parallel transport basis as follows: $$H_{\text{kin}} = \epsilon_{\bm{k}} \left[ c^{\dagger}_{\bm{k},1,K} \, c^{\phantom{\dagger}}_{\bm{k},2,K} - c^{\dagger}_{\bm{k},2,K'} \, c^{\phantom{\dagger}}_{\bm{k},1,K'} + \text{h.c.} \right] = \begin{pmatrix} c^{\dagger}_{\bm{k},1,K} & - c^{\dagger}_{\bm{k},2,K'} \end{pmatrix} \begin{pmatrix} c^{\phantom{\dagger}}_{\bm{k},2,K} \\ c^{\phantom{\dagger}}_{\bm{k},1,K'} \end{pmatrix}  + \text{h.c.}, $$ suppressing the spin indices. The creation operators in the row vector correspond to one Chern sector and the annihilation operators in the column vector belong to the opposite Chern sector; this shows that the two unitary matrices acting on the two separate Chern sectors need to be related so that the kinetic term remains invariant as well. In other words, if, for example, the $2\times2$ unitary $U$ is used for the $C=+1$ sector, $ U \mu^z$ ($\mu^z$ is the Pauli matrix acting on the above doublets of fermion operators) should be used for the $C=-1$ sector. This reduces the symmetry group to U(4) when spin is also included.
    \item $\eta \neq 0$: away from the chiral limit, apart form the bilinears 
    $\begin{pmatrix} c^{\dagger}_{\bm{k}_1,1,K} & c^{\dagger}_{\bm{k}_1,2,K'} \end{pmatrix} \begin{pmatrix} c^{\phantom{\dagger}}_{\bm{k}_4,1,K} \\ c^{\phantom{\dagger}}_{\bm{k}_4,2,K'} \end{pmatrix}$ 
    and 
    $\begin{pmatrix} c^{\dagger}_{\bm{k}_1,2,K} & c^{\dagger}_{\bm{k}_1,1,K'} \end{pmatrix} \begin{pmatrix} c^{\phantom{\dagger}}_{\bm{k}_4,2,K} \\ c^{\phantom{\dagger}}_{\bm{k}_4,1,K'} \end{pmatrix}$,
    other combinations also appear in the four Fermi interaction; these terms could be written as:
    $\begin{pmatrix} c^{\dagger}_{\bm{k}_1,1,K} & c^{\dagger}_{\bm{k}_1,2,K'} \end{pmatrix} \begin{pmatrix} c^{\phantom{\dagger}}_{\bm{k}_4,2,K} \\ c^{\phantom{\dagger}}_{\bm{k}_4,1,K'} \end{pmatrix}$ 
    and 
    $\begin{pmatrix} c^{\dagger}_{\bm{k}_1,2,K} & c^{\dagger}_{\bm{k}_1,1,K'} \end{pmatrix}  \begin{pmatrix} c^{\phantom{\dagger}}_{\bm{k}_4,1,K} \\ c^{\phantom{\dagger}}_{\bm{k}_4,2,K'} \end{pmatrix}$. This again means that the unitaries in the two Chern sectors should be related, and in fact identical so that these new terms also remain invariant. This results in the symmetry group U(4) with spin included; this last symmetry of the interaction term away from the chiral limit does not survive when the noninteracting terms of the Hamiltonian are considered.
\end{itemize}

\end{itemize}

\section{Particle-hole symmetry between $\nu = +3$ and $\nu = -3$}\label{app:ph_hwf}

In this Appendix, we discuss how the particle hole symmetry of the CM is displayed in the way the many body states are transformed between the two fillings $\nu = \pm3$. We should note that for general $\eta$, we have an approximate particle hole symmetry which needs a $k_x \to -k_x$ transformation as well. As was discussed in the main text, this particle hole symmetry is broken in our numerical results for the first study, i.e.~when the HF zero point is taken at $\nu = -4$, or in other words when only the HWF basis hoppings along with the interaction between particles in the active bands are kept. It is broken even in the limit of $\eta = 0$, i.e.~the chiral model regardless of the value of $\Delta$. We will furthermore argue that had we started with a model where the zero point of the HF is at $\nu = +4$, we would have gotten the particle hole transformed version of the same model; in this model holes will play the role of electrons. Finally we will sketch how the particle hole symmetry is retained in the projected model of our second study.

We consider the model of our first study in the chiral limit for simplicity. The chiral symmetry of the model in this limit indicates that each state with an energy $E(\bm{k})$ has a counterpart with the same $\bm{k}$ value that but opposite energy $-E(\bm{k})$. Note again that this is different from the particle hole symmetry we discussed above (the latter is present with an approximation of neglecting the rotation of sublattice matrices); we only consider the chiral limit in the following but very similar reasoning can be done for the particle hole symmetry at general $\eta$. The two states with energies $\pm E(\bm{k})$ could be written in terms of each other as $\left| \psi_{\bm{k} , 1 , \xi} \right \rangle = i \; \sigma^z \left| \psi_{\bm{k} , 2, \xi} \right \rangle$, where the indices $1,2$ correspond to states within the two active bands. Thus, it is easy to form sublattice polarized states:
\begin{equation}\label{eq:HWFs_eta_0_appendix}
	\tilde{c}^{\dagger}_{\bm{k},m,\xi} = \frac{e^{i \phi_{k_x;k_y,m,\xi}}}{\sqrt{2}} \left[ c^{\dagger}_{\bm{k},1,\xi} + (-1)^{m} i \; c^{\dagger}_{\bm{k},2,\xi} \right]. 
\end{equation}
It has been argued in the main text and the Appendix that the states in the parallel transport basis also have such a form and thus we take $\tilde{c}^{\dagger}_{\bm{k},\xi,m}$ to be the creation operator in the parallel transport basis. One can get the maximally localized HWFs by doing a Wannier transform:
\begin{equation}
	c^{\dagger}_{  k_x ; y , m , \xi} = \frac{1}{N_y} \sum_{k_y} e^{-i k_y y} \, \tilde{c}^{\dagger}_{\bm{k},m,\xi}.
\end{equation}
Note that on the left hand side, i.e.~Fermi operators in the HWF basis we are not using $\tilde{\cdot}$ signs anymore. We will also drop the subscript of $k_x$. The kinetic term of the Hamiltonian in terms of these states reads:
\begin{equation}
\begin{aligned}
	\left. H_{\text{kin}} \right|_{\Delta = 0} &= \frac{1}{N_x N_y} \sum_{\xi,\bm{k}} \epsilon_{\bm{k},\xi} \,   \left[e^{i \varphi_{k,k_y,\xi}} \ \tilde{c}^{\dagger}_{\bm{k},1,\xi}  \, \tilde{c}^{\phantom{\dagger}}_{\bm{k},2,\xi} + e^{-i \varphi_{k,k_y,\xi}} \ \tilde{c}^{\dagger}_{\bm{k},2,\xi}  \, \tilde{c}^{\phantom{\dagger}}_{\bm{k},1,\xi}  \right] \\
	&= \frac{1}{N_x} \sum_{\xi,k,yy'} \,   \left[ t^{\, y'-y  ,  k , \xi} \ c^{\dagger}_{k,y',1,\xi}  \, c^{\phantom{\dagger}}_{k,y,2,\xi} + \left(t^{\, y'-y  ,  k , \xi}\right)^* \ c^{\dagger}_{k,y,2,\xi}  \, c^{\phantom{\dagger}}_{k,y',1,\xi}  \right],
	\end{aligned}
\end{equation}
where $\varphi_{k,k_y,\xi} = \phi_{k,k_y,2,\xi} - \phi_{k,k_y,1,\xi}$, and the hopping parameter reads $t^{\, y'-y  ,  k , \xi} = \frac{1}{N_y} \sum_{k_y} e^{ik_y(y'-y)}   \left[ \epsilon_{\bm{k},\xi} \ e^{i \varphi_{k,k_y,\xi}} \right]$, and the subscript of $k_x$ is not shown from here on. When the sublattice potential term is also present, we have argued above that since the term $\sigma^z \Delta$ keeps the active bands subspace intact at each $\bm{k}$, the states shown in \eqref{eq:HWFs_eta_0_appendix} still form the HWF basis; it is straightforward to work out the sublattice potential form as well, since HWF basis is sublattice polarized:
\begin{equation}
	H_{\Delta} = \frac{1}{N_x} \Delta  \sum_{\xi,k,y} \left[ \ c^{\dagger}_{k,y,1,\xi}  \, c^{\phantom{\dagger}}_{k,y,1,\xi} - c^{\dagger}_{k,y,2,\xi}  \, c^{\phantom{\dagger}}_{k,y,2,\xi}  \right].
\end{equation}
We also note that $\phi_{k,k_y,2,\xi} = - \phi_{k,k_y,1,\xi}$, regardless of the value of $\Delta$.

Now, it is straightforward to check that the terms in $H_{\text{kin}}$, including the $\Delta$ term, have the same form in terms of $d$ operators as that in terms of $c$ operators, where they are defined as in the following particle hole transformations:
\begin{equation}\label{eq:ph_definition_c_cdagger}
    c^{\dagger}_{k,y,1,\xi} = d^{\phantom{\dagger}}_{k,-y,2,\xi} , \quad c^{\dagger}_{k,y,2,\xi} = - d^{\phantom{\dagger}}_{k,-y,1,\xi},
\end{equation}
This means that $H_{\text{kin}}$ is particle hole symmetric with the above prescription. Spin indices could be trivially added to the above terms.

\vspace{0.5cm}

\begin{figure}[!t]
	\centering
	\subfigure[]{\includegraphics[height=0.225\textwidth]{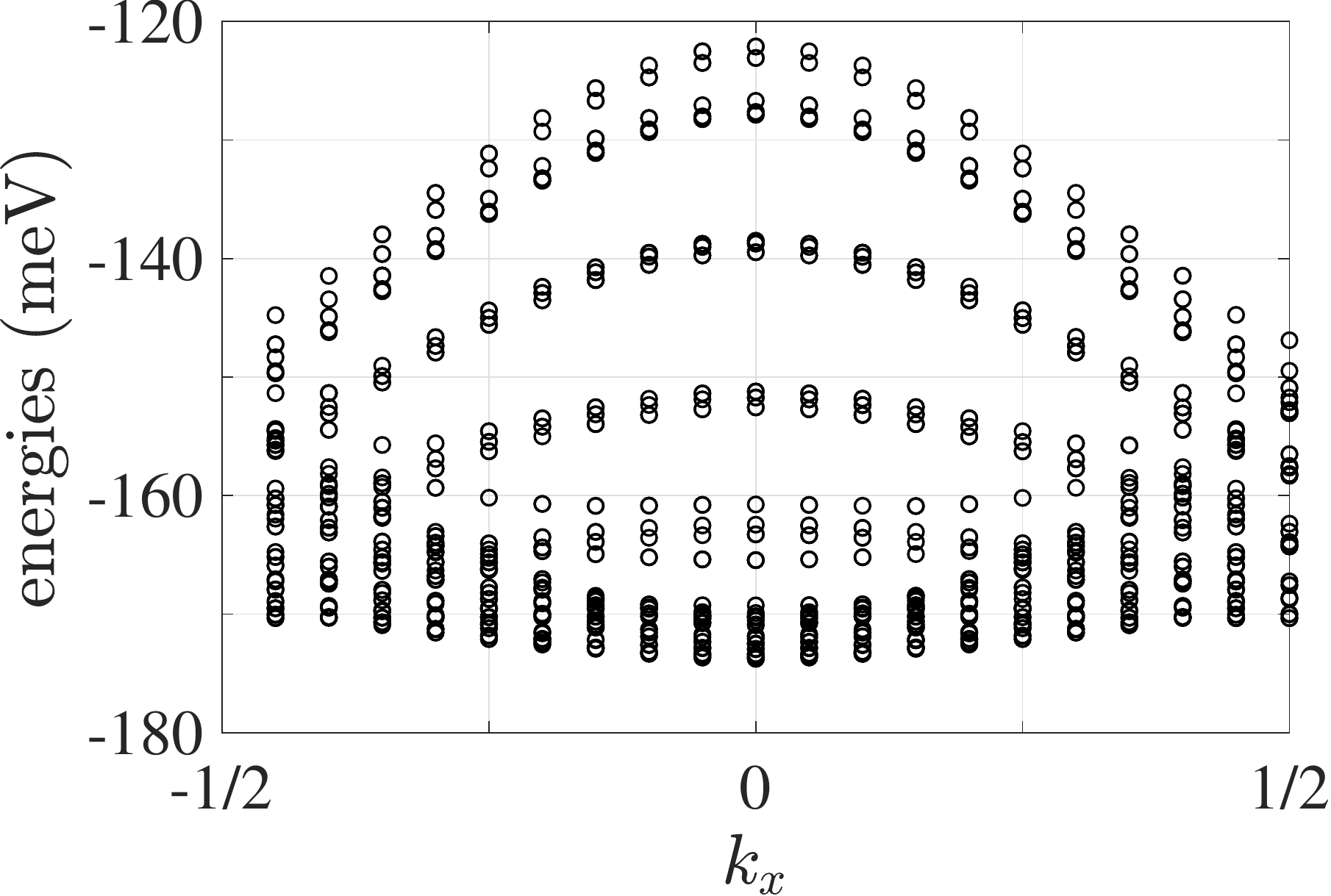}}
	\qquad
	\subfigure[]{\includegraphics[height=0.225\textwidth]{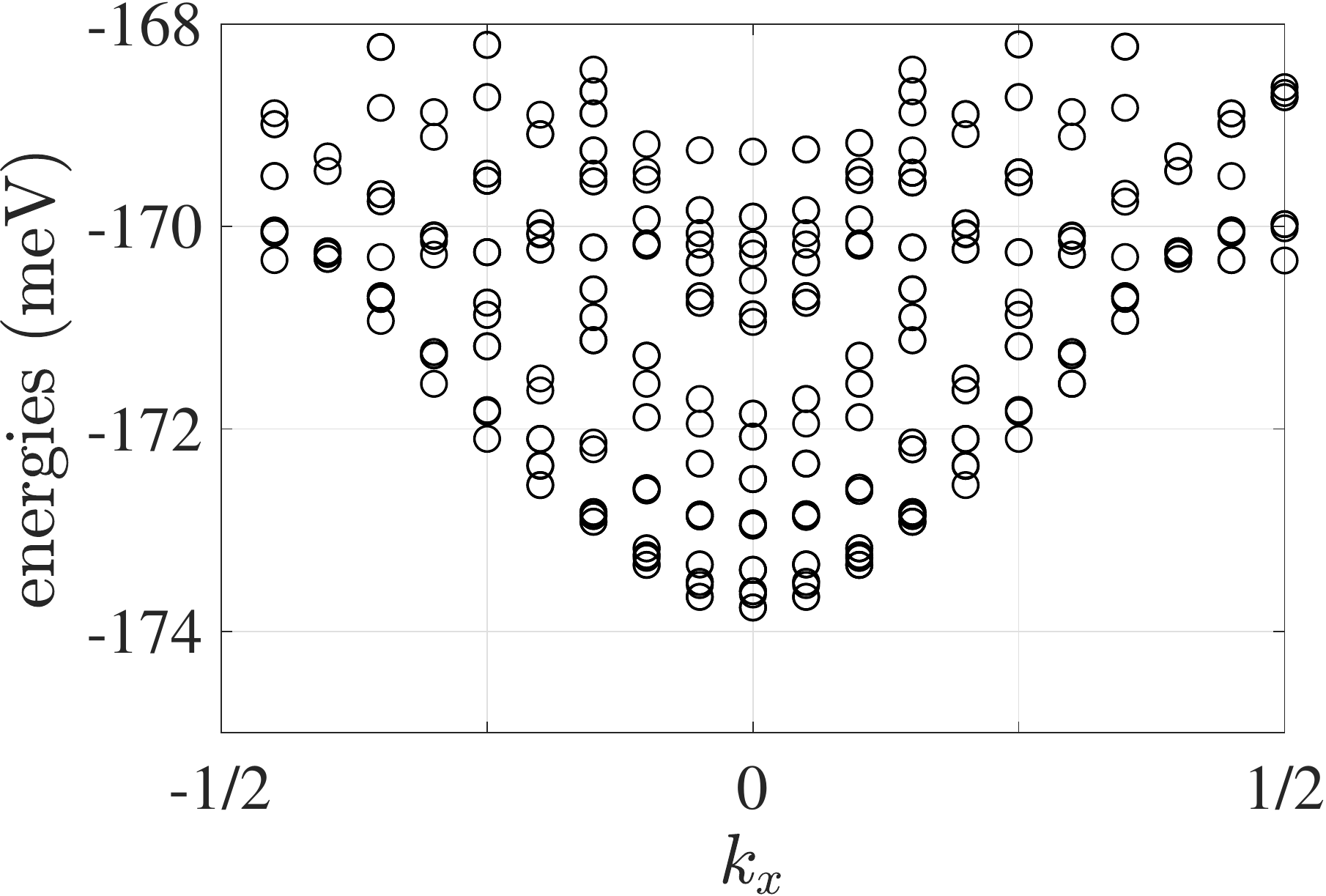}}
	\caption{\footnotesize Eigenvalues of the single hole potential shown in \eqref{eq:hartree_factors}. The following set of paramters has been used: $\eta = 0$, $\alpha = 0.58$, $\Delta = 1.9 \text{meV}$, $\ell_\xi = 0.1 a_M$, $g_{\text{int}} = 0.1$. Lower energies are available for smaller $k_x$ values. A magnified view of smallest energies is shown in panel (b).
	}
	\label{fig:ph_potential}
\end{figure}

One should furthermore consider the interaction term;  the interaction in general can be written as follows:
 \begin{equation}
 \begin{aligned}
 		H_{\text{int}} =  \frac12 \sum_{k\text{'s} ,\, y \text{'s}} \ \sum_{\xi\xi' s s'} \  \sum_{mm'} \ 
 		 V_{k_1y_1\xi m; \, k_2y_2\xi' m' ; \, k_3y_3\xi' m' ; \, k_4y_4\xi m}   \ \left[ c^{\dagger}_{k_1,y_1,m,\xi,s}  \,  c^{\dagger}_{k_2,y_2 ,m',\xi',s'}  \, c^{\phantom{\dagger}}_{k_3,y_3 ,m',\xi',s'} \, c^{\phantom{\dagger}}_{k_4,y_4 ,m,\xi,s} \right].
 \end{aligned}
 \end{equation}
Since, $\langle  \bm{r} ,\sigma\tau  \left| k_1y_1 m \xi \right.\rangle = \langle  -\bm{r} ,\bar{\sigma}\tau  \left| k_1 (-y_1) \bar{m} \xi  \right\rangle^*$, the interaction terms have the following relations between themselves:
$$ V_{k_1y_1\xi m; \, k_2y_2\xi' m' ; \, k_3y_3\xi' m' ; \, k_4y_4\xi m}  =  V_{k_4 (-y_4) \xi \bar{m}; \, k_3 (-y_3) \xi' \bar{m}' ; \, k_2 (-y_2) \xi' \bar{m}' ; \, k_1 (-y_1) \xi \bar{m}} .$$
The interaction thus takes the following form in terms of the $d$ operators:
\begin{equation}\label{eq:hamiltonian_in_terms_hole_ops}
\begin{aligned}
	H_{\text{int}}  =&  \frac12 \sum_{1234} V_{1,2,3,4} \ d^{\dagger}_{1} \, d^{\dagger}_{2}  \,   d^{\phantom{\dagger}}_{3}  \, d^{\phantom{\dagger}}_{4}  \\
	 & +   \sum_{123} d^{\dagger}_{2}  \,   d^{\phantom{\dagger}}_{3}  \left( -  V_{1231} + V_{1213} \right) \\
	 & + \frac12 \sum_{12} \left(  V_{1221}  - V_{1212} \right) ,
\end{aligned}
\end{equation}
where for simplicity a change of notation has been made $1 \equiv (k_1,y_1,\xi,s,m)$, and so forth. The first term above is identical in form to the original interaction Hamiltonian in terms of $c$ operators. However, there are terms quadratic in $d$, the \textit{single hole terms}, that were not present in the original Hamiltonian. Note that the terms on the third row are constant. The terms on the second row, on the other hand, turn out to be $k_x$-dependent and thus impose a single hole potential; this is the origin of the particle hole asymmetry between the fillings $\pm 3$, as was discussed in the main text.
Note that unlike this situation, in the usual Hubbard model with a single band, nearest neighbor hopping and constant on-site interaction for example, the analogue of this term is just a redefinition of the chemical potential.

The single hole potential introduced above has the following explicit form:
\begin{equation}\label{eq:hartree_factors}
\begin{aligned}
	 \sum_{k_2  y_2 y_3} & \sum_{m_2  \xi_2 s_2} d^{\dagger}_{k_2 , y_2 , m_2 , \xi_2 , s_2}  \,   d^{\phantom{\dagger}}_{k_2 , y_3 , m_2 , \xi_2 , s_2}\\
	& \sum_{k_1 y_1 ; m_1 \xi_1} \bigg[ - 2 V_{k_1 y_1 \xi_1 m_1; \, k_2 y_2 \xi_2 m_2 ; \, k_2 y_3 \xi_2 m_2 ; \, k_1 y_1 \xi_1 m_1}    + \delta_{\xi_2\xi_1} \delta_{m_2m_1} \ V_{k_1 y_1 \xi_1 m_1; \, k_2 y_2 \xi_1 m_1 ; \, k_1 y_1 \xi_1 m_1 ; \, k_2 y_3 \xi_1 m_1} \bigg],
\end{aligned}
\end{equation}
This single hole potential has been calculated numerically for a special case and its eigenvalues are formed, see Fig.\ref{fig:ph_potential}. One can observe that hole states with $k_x$ closer to $0$ are preferred.

We briefly mention here what form of a particle hole transformation should be used, instead of \eqref{eq:ph_definition_c_cdagger}, when $\eta \neq 0 $ which mean that there is no chiral symmetry in the model. Generically and regardless of the value of $\Delta$, the following transformation could be used:
\begin{equation}\label{eq:general_ph_Delta_nonzero}
	\tilde{c}^{\dagger}_{\bm{k},1,\xi} = \tilde{d}^{\phantom{\dagger}}_{-\tilde{\bm{k}},2,\bar{\xi}} , \quad \tilde{c}^{\dagger}_{\bm{k},2,\xi} = - \tilde{d}^{\phantom{\dagger}}_{-\tilde{\bm{k}},1,\bar{\xi}},
\end{equation}
where for $\bm{k}=(k_x,k_y)$, we define $\tilde{\bm{k}} = (-k_x,k_y)$. Note that for the sake of clarity we have expressed the particle hole transformation for creation and annihilation operators in the parallel transport basis, i.e.~before the hybrid Wannier transformation is performed. It is straightforward to re peat the manipulations detailed above also with this transformation. If $\Delta = 0$, the $C_2\mathcal{T}$ is present and one can use a particle hole transformation that works within each valley:
\begin{equation}\label{eq:general_ph_Delta_zero}
	\tilde{c}^{\dagger}_{\bm{k},1,\xi} = \tilde{d}^{\phantom{\dagger}}_{\tilde{\bm{k}},2,\xi} , \quad \tilde{c}^{\dagger}_{\bm{k},2,\xi} = - \tilde{d}^{\phantom{\dagger}}_{\tilde{\bm{k}},1,\xi}.
\end{equation}
It is worthwhile to note that \eqref{eq:general_ph_Delta_nonzero} preserves the Chern number of the band, while \eqref{eq:general_ph_Delta_zero} takes it to the opposite value. In Fig.~\ref{fig:ph_hf_spectra} of the main text we have used the latter transformation since $C_2\mathcal{T}$ is present.

It is simple now to see how one could obtain a model with its HF zero point at $\nu = +4$; by requiring the second row in Eq.~\eqref{eq:hamiltonian_in_terms_hole_ops} to be cancelled by the terms in $H_{\text{MF},0}$. Note that this will result in a Hamiltonian which is identical to the one we used in our first study, except that the electrons are replaced by holes.
It is also easy at this point to check that the model with its zero point at the CNP is particle hole symmetric. This happens due to the particular form that $H_{\text{MF},0}$ takes for this choice, i.e.~Eq.~\eqref{eq:H_MF_second_approach}; it is straightforward to check that the sum of $H_{\text{MF},0}$ with the terms on the second row of Eq.~\eqref{eq:hamiltonian_in_terms_hole_ops} takes the form of $H_{\text{MF},0}$ again but particle hole transformed.

\section{Comparison with other Hartree Fock studies}\label{app:comparison_hf}
In this Appendix, we compare our approach and results on the HF stability of QAHE with other recent HF studies, namely Refs.~\onlinecite{liu2019nematic,liu2019spontaneous,xie2020nature,bultinck2019ground}. We first summarize our results: our numerical analysis shows that with the physical choice of $\eta \approx 0.8$, we observe a robust QAHE for $\nu=-3$ and $\nu=+3$, if we set the zero point of our HF approach to be at $\nu=-4$ and $\nu=+4$ respectively. This QAHE is a consequence of valley, spin and band (in the HWF basis) polarization in the HF solutions. On the contrary, if the HF zero point is taken at the CNP, there is a particle hole symmetry between the many body states found at $\nu=\pm 3$; we only observe QAHE in small windows of parameters in either of these two filling factors for the choice of $\eta = 0.8$.
Based on these observations and following a phenomenological argument, we expect the model with the HF zero point set at $\nu=+4$ to be most relevant to physics seen in TBG samples exhibiting QAHE. In the following, we compare the results of this model with those presented in some of the recent related HF studies.

We start with Ref.~\onlinecite{xie2020nature}, where a HF study is carried out keeping the remote bands as dynamical in the analysis. Furthermore, the zero point of the HF Hamiltonian is taken at the CNP of decoupled monolayer graphene sheets. These authors have considered several filling factors, for example at CNP, they observe an interaction induced gap corresponding to a $C_2\mathcal{T}$ broken phase for large enough interaction. On the other hand, in the insulators they obtain at the fillings $\nu=\pm 1$ and $\nu = \pm 3$, $C_2\mathcal{T}$ is not necessarily broken and thus the many body states at these fillings do not automatically show QAHE. This is in contrast to our findings outlined above where an insulator exhibiting QAHE could be observed at one of these two fillings depending on the choice of the HF zero point.

We next turn to Ref.~\onlinecite{liu2019nematic}, where a HF study has been implemented taking only the active bands as dynamical. The zero point of their HF is set at the CNP of the active bands, and this makes their model similar to the one in one of our studies. The focus of this work is on the CNP and they report observing a variety of different symmetry broken insulating states in their numerical results, including $C_2\mathcal{T}$ broken, spin/valley polarized, etc. Ref.~\onlinecite{bultinck2019ground}, on the other hand, deals with the full set of moir\'e bands in the HF analysis, but with the main focus on the CNP also. Interestingly, the $U(4)\times U(4)$ symmetry of the chiral model ($\eta=0$) discussed in this work can also be seen in the HWF basis (and also the parallel transport basis) as discussed in the present paper; for general $\eta$, when $C_2\mathcal{T}$ is present, an interaction-only model consisting of active bands only displays a $U(4) \times U(4)$ symmetry (see Appendix \ref{app:hamiltonian_hwf}).

Finally, we consider Ref.~\onlinecite{liu2019spontaneous}, where a HF study taking all bands into account has been presented. The authors consider several filling factors, and in particular, they are able to see a QAHE at $\nu = \pm 3$; the presence of a significant sublattice potential is crucial for the QAHE to materialize. This is in contrast to the present work, where the presence of a sublattice potential can make the QAHE stronger, but it is not necessary for the occurrence of the required flavor polarization. Within our study, we observed that a larger interaction strength could compensate for the absence of the sublattice potential.

\end{document}